\pdfoutput=1
\documentclass[]{jfm}
\usepackage{graphicx}
\usepackage{newtxtext}
\usepackage{newtxmath}
\usepackage{natbib}
\usepackage{hyperref}
\hypersetup{
    colorlinks = true,
    urlcolor   = blue,
    citecolor  = black,
}

\newcommand{\RomanNumeralCaps}[1]
\linenumbers

\usepackage{graphicx}
\usepackage{natbib}
\usepackage{natmove}
\usepackage[labelfont=bf,textfont=normalfont,singlelinecheck=on,justification=raggedright]{caption}
\usepackage[labelfont=bf,textfont=normalfont,singlelinecheck=on,justification=justified]{subcaption}

\usepackage[section]{placeins}
\usepackage{color}

\usepackage{psfrag}               
\usepackage{lipsum}
\usepackage{amssymb}
\usepackage{gensymb}
\usepackage{latexsym}

\usepackage{epsfig}
\usepackage{pstricks}
\usepackage{afterpage}
\usepackage{nicefrac}
\usepackage{textcomp}

\usepackage{tikz}
\usepackage{hyperref}

\graphicspath{{images/}}

\title{Analysis of the flow of granular materials through a screw conveyor}

\author{Aashish Kumar Gupta\aff{1}
  \corresp{\email{aashishgupta@iisc.ac.in}}
  \and
  Prabhu R. Nott\aff{1}
 }

\affiliation{\aff{1}Department of Chemical Engineering, Indian Institute of Science, Bangalore, India}
\date{June 2020}

\begin{document}
\maketitle

\begin{abstract}
	Screw conveyors are widely employed in industry for the bulk transport of particulate materials. Several studies have attempted to correlate the discharge rate with the rotation speed of the screw via experiments and particle dynamics simulations. However, a detailed mechanical model that would assist in the optimal design of screw conveyors has not been attempted. In this study, we first construct a simple model that assumes the entire granular medium to move as a rigid body sliding along the surfaces of the screw and barrel. By enforcing the balances of linear and angular momentum to a suitably chosen continuum element, we show that under certain limiting conditions, the discharge rate for a given angular velocity and screw geometry can be obtained. Further, we show that the discharge can be maximized by setting the ratio of the pitch to barrel diameter to a particular value. We then study the detailed flow within the conveyor using the Discrete Element Method, which reveals that a significant fraction of the material exhibits solid body motion, in agreement with the simple model. We assess the effect of relaxing the limiting conditions employed in the model, thereby determining the connection between the friction at the walls and the kinematics of flow. Finally, the trend in variation of average axial velocity with the pitch to barrel diameter ratio in the presence of gravity is compared with that from our model and simulations in the absence of gravity. We observe that both the trends are qualitatively the same indicating that the dependence of average axial velocity on the geometry of the conveyor is not altered by the introduction of gravity.
\end{abstract}

\section{Introduction}
\noindent Screw conveyors have been in use since the ancient Greek and Egyptian civilizations to draw water against gravity \citep{chakarborthy2014product}. In current times, they are employed extensively for conveying and processing of particulate materials in the pharmaceutical, food processing and mineral industries. Despite their considerable importance in industry, a fundamental understanding of the performance of screw conveyors based on continuum mechanical principles is lacking.  The reasons for this are: the geometry of the flow is complex, robust continuum models for dense particulate materials are still in a stage of development, and the materials conveyed vary widely in form and physical properties (ranging from  dry grains to liquid-saturated pastes, and fine powders to coarse granules).  As a result, most previous attempts to study the problem have used particle-based computational methods to understand the flow rate as a function of rotational speed of the screw, the volumetric fill level of the bulk material, inclination of the screw with horizontal plane \citep{owen2009prediction}, magnitude of forces between cohesive grain particles \citep{hou2014study} and so on. The complex nature of flow inside a screw conveyor has resisted attempts to theoretically model the flow. However, an expression for discharge was provided by \cite{roberts1999influence} by drawing analogy with a plunger scraping material through a cylindrical shell. An analysis of flow was also attempted by \cite{wu1978transport} by reducing the geometry to a rectangular channel of an equivalent inclination. In practical design, a rough estimate of the performance is based on an empirical equation from the database of practical machines\citep{colijn1985mechanical,woodcockmason}. In this work, we aim to provide insight into the dependence of flow rate predominantly on the geometry of the screw conveyor.\\

\noindent The current work aims to develop a model based purely on mechanics to relate the discharge to the rotation speed of screw, without simplifying the complexity of the system geometry. We will be working with a horizontal screw conveyor where the rotation speed of the inner shaft is small enough that the flow is in the slow flow regime. The effect of gravity is ignored so that we may understand the effects of the geometry and friction on the flow. The analysis is described in \S 2, where it is assumed that the granular material moves as a solid plug that slides along the surface of the screw. A differential element of the plug is suitably chosen on which the momentum balances are enforced in order to arrive at an expression for the rate of discharge. We show that without knowledge of the stresses on the various surfaces, a solution of discharge is only admissible in the limiting case of friction vanishing on the screw surface. Despite this simplification, the model provides good insight into the mechanism of conveyance. An interesting feature of the prediction is that the flow rate is maximum at a particular value of the ratio of the screw pitch to barrel diameter.\\\\
\noindent Furthermore, we employ particle dynamics simulations using the Discrete Element Method (DEM), where we impose a finite coefficient of friction on the screw and the barrel surfaces, and study scenarios with and without gravity. An interesting result is that the variation of the flow rate  with the ratio of screw pitch to barrel diameter is similar in these simulations irrespective of the situation of gravity , and also align with the prediction of our simple model. These results suggest a tentative range for the pitch to barrel diameter ratio to be maintained while designing for maximum throughput.

\section{Theoretical Modelling}
A simplistic model proposed by \cite{roberts1999influence} to relate the flow rate through a screw conveyor to its rotation speed, replaces the complex geometry of the conveyor with a plunger. The screw flight is reduced to its projection on a plane perpendicular to the axis of the screw, viz. an annular disc. This disk slides on the screw shaft pushing the material inside the barrel by a distance equal to the pitch, $p$, of the screw in a duration corresponding to the time period of rotation of the screw. The corresponding equation for discharge, $Q$, is as follows.\\
\begin{equation}
Q=\nu \pi (r_b^2-r_s^2)pN
\label{n_vs_p}
\end{equation}\\
where $\nu$ is the solid fraction, $N$ is the number of rotation of the screw per second, and $r_b$ and $r_s$ are the radii of the barrel and the screw shaft respectively.\\
\\
It is to be noted that, owing to the rotation of the screw inside a screw conveyor, the granular material should posses an azimuthal velocity too in addition to the axial component. However, the mere pushing action of the annular disk proposed in this model cannot impart any rotation to the granular material. Additionally, the flux of material as suggested by Eq.\ref{n_vs_p}  is a monotonic function of the pitch of the screw. However, with an extremely large pitch for a given shaft and barrel radii, as the flights become aligned to the axis of the screw, the screw rotation should only impart an azimuthal velocity and the axial flux should be zero. \\

We now discuss our model that relates discharge to the geometry of the screw and its rotation speed. We choose a geometry where there is no clearance between the screw blade and the barrel surface, and the entire available space is filled with grains. As we are operating at low screw rotation speeds, we neglect centrifugal effects on the granular material, and the effect of gravity. The most crucial assumption is that of the entire granular medium behaving as a solid plug that slides inside the helical channel as a response to the screw rotation. The screw flight surface is described mathematically and three primary direction vectors necessary for the analysis are arrived at. A crude expression for the rate of discharge is also obtained. Later, balances of linear and angular momentum are enforced on a suitably chosen differential element to obtain an exact expression for discharge in the limit of vanishing friction on the screw surface.\\

\begin{figure}
	\begin{minipage}{.5\linewidth}
		\centering
		\begin{subfigure}{\linewidth}
			\includegraphics[width=\linewidth]{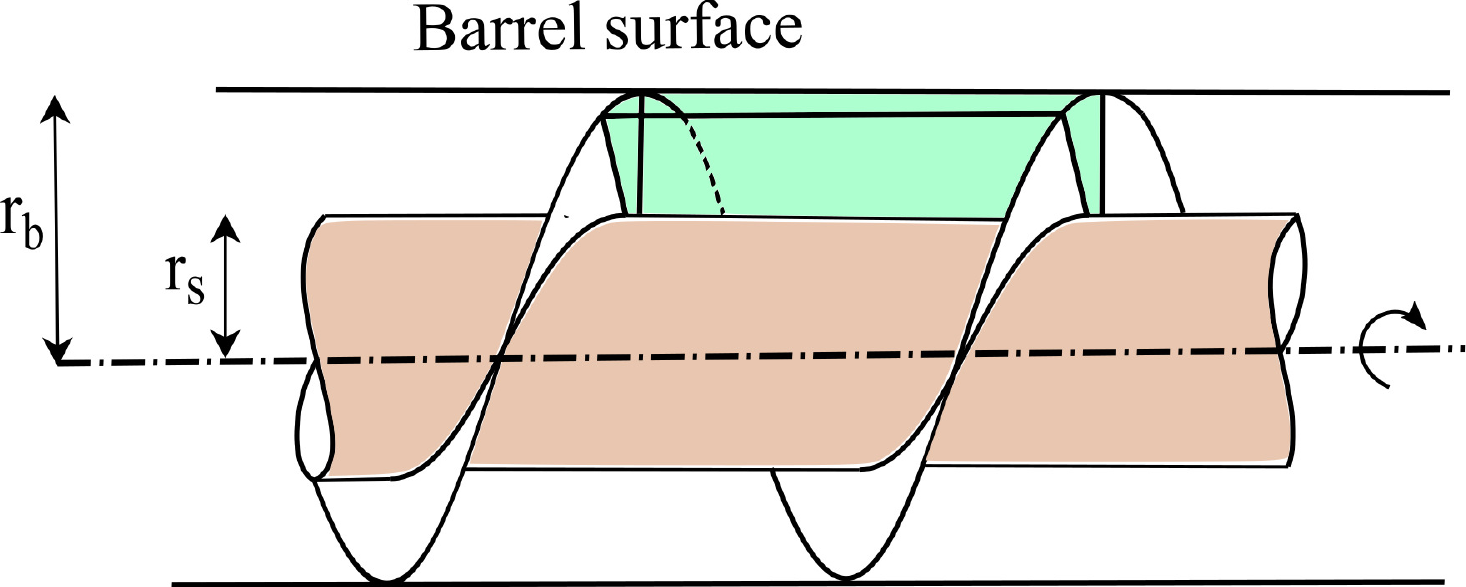}
			\caption{}
			\label{elem_loc_a}
		\end{subfigure}\\[1ex]
	\end{minipage}%
	\begin{minipage}{.5\linewidth}
		\centering
		\begin{subfigure}{\linewidth}
			\includegraphics[width=\linewidth]{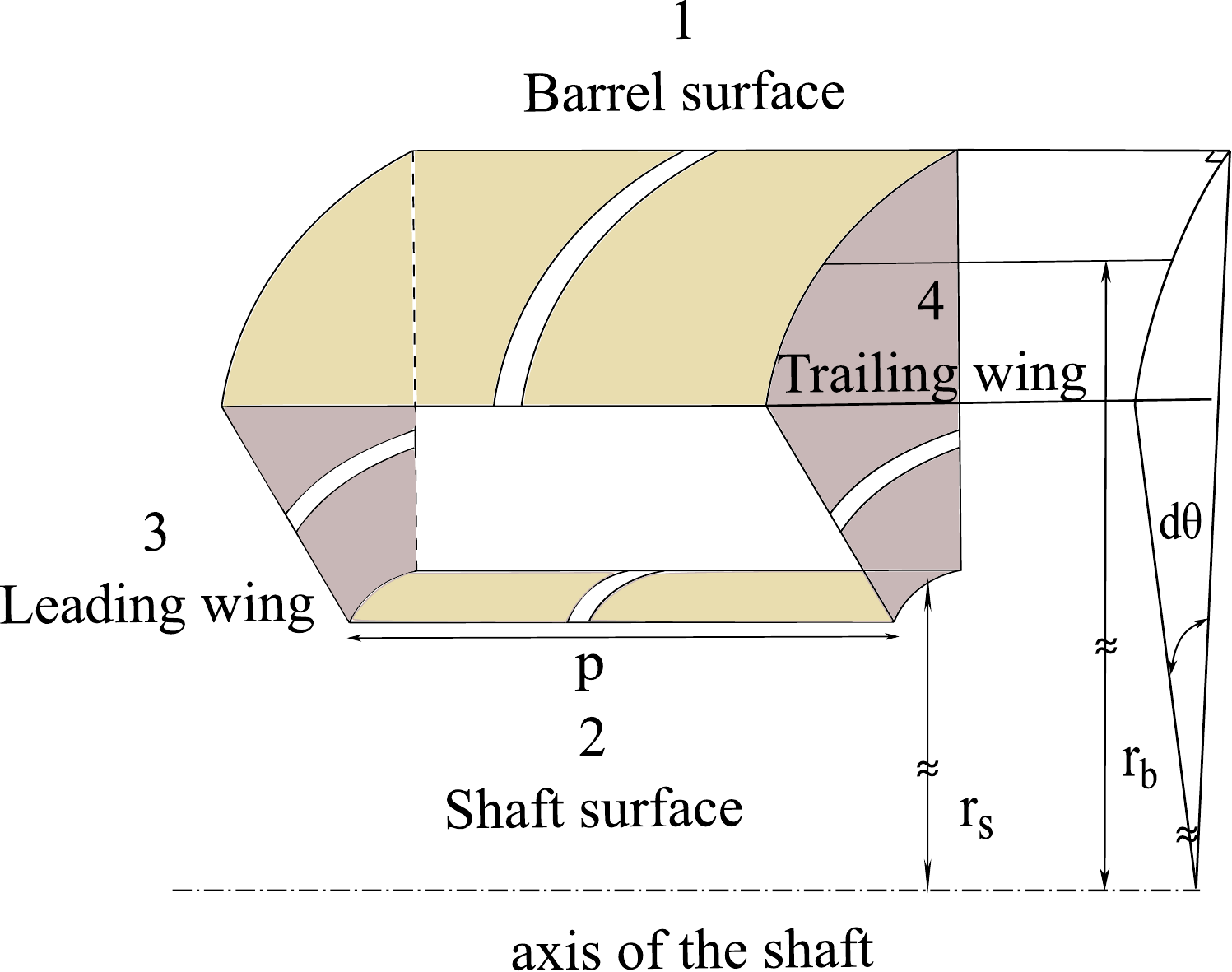}
			\caption{}
			\label{elem_loc_b}
		\end{subfigure}\\[1ex]
	\end{minipage}
	
	\caption{(a) location of the element (highlighted in green) chosen for the momentum balance , (b) an enlarged view of the element under consideration}
	\label{element}
\end{figure}

The assumption that the material inside the conveyor moves as a solid plug makes it possible for us to consider an element which extends up to the screw surfaces and the barrel, as opposed to an infinitesimally small element. Consider the wedge-shaped element of angular width $d\theta$, shown in Figure \ref{element}, whose left and right boundaries are the leading and trailing flights of the screw, and top and bottom boundaries are the surfaces of the barrel and shaft, respectively. The advantage of such an element is that the forces acting on four of its faces are the traction on the screw and the barrel surfaces themselves. However, the traction in general might vary over these surfaces. So, we consider sub-elemental areas on each of these surfaces, shown as white strips in the figure, over which the traction is assumed to remain constant. As the screw rotates, this element slides in the helical channel and is ultimately ejected out.\\

The screw-flight surface is mathematically described by a family of helices of a given pitch. Owing to the differences in the circumferential distance that the helices at different radial locations cover while traversing an axial distance equal to their pitch, the helix angle $\alpha$, which is the angle between the local tangent to the helix and the azimuthal direction ,varies with $r$. For a helix situated on the flight at a distance $r$ from the axis of the screw, the helix angle, $\alpha(r)$, is related to the pitch , $p$, and $r$ as follows.
\begin{equation}
\tan{\alpha(r)}=\frac{p}{2\pi r}
\label{angle}
\end{equation}
\\
\FloatBarrier
\noindent Let the position vector of a point on the helix that corresponds to a rotation $\theta$ radians from a reference axis be as follows.
\begin{equation}
{\overrightarrow{P_h}(r,\theta)= r \hat{e_r}- \left(\frac{p\theta}{2\pi}\right)\hat{e_z}}
\end{equation}

\noindent A vector tangent to the helix at that point can be calculated simply by taking the partial derivative of the position vector with respect to the angle $\theta$. Hence, the expression for the unit tangential vector becomes
\begin{equation}
\begin{aligned}
\hat{t_h}(r,\theta)&=\left(\frac{\partial \overrightarrow{P_h}(r,\theta)}{\partial \theta}\right)/{\left|\frac{\partial \overrightarrow{P_h}(r,\theta)}{\partial \theta}\right|}\\
&=\cos\alpha \, \hat{e_\theta}-\sin\alpha \, \hat{e_z}
\end{aligned}
\label{t_h}
\end{equation}

\vspace{10 pt}
\noindent Similarly, the unit vector which at right angle from
$\hat{t_h}(r,\theta)$ in the counter-clockwise direction, and which lies on the surface of the screw-flight is given as follows.
\begin{equation}
\begin{aligned}
\hat{s_h}(r,\theta)&={\left(\frac{\partial \overrightarrow{P_h}(r,\theta)}{\partial r}\right)}/{\left|\frac{\partial \overrightarrow{P_h}(r,\theta)}{\partial r}\right|}\\
&=\hat{e_r}
\end{aligned}
\label{s_h}
\end{equation}	

\vspace{10 pt}
\noindent The unit vector that is perpendicular to the local flight surface, $\hat{n_h}(r,\theta)$, can be obtained by taking a cross-product of the vectors $\hat{t_h}(r,\theta)$ and $\hat{s_h}(r,\theta)$ that lie in the plane of the flight. 
\begin{equation}
\hat{n_h}(r,\theta)=\hat{t_h}\times \hat{s_h}
\end{equation}	
Using Eq.\ref{t_h} and Eq.\ref{s_h} in the above equation, we get
\begin{equation}
{\hat{n_h}(r,\theta)=-\sin\alpha \, \hat{e_\theta}-\cos\alpha \, \hat{e_z}}
\end{equation}	

\begin{figure}
	\begin{minipage}{.5\linewidth}
		\centering
		\begin{subfigure}{0.9\linewidth}
			\includegraphics[width=\linewidth]{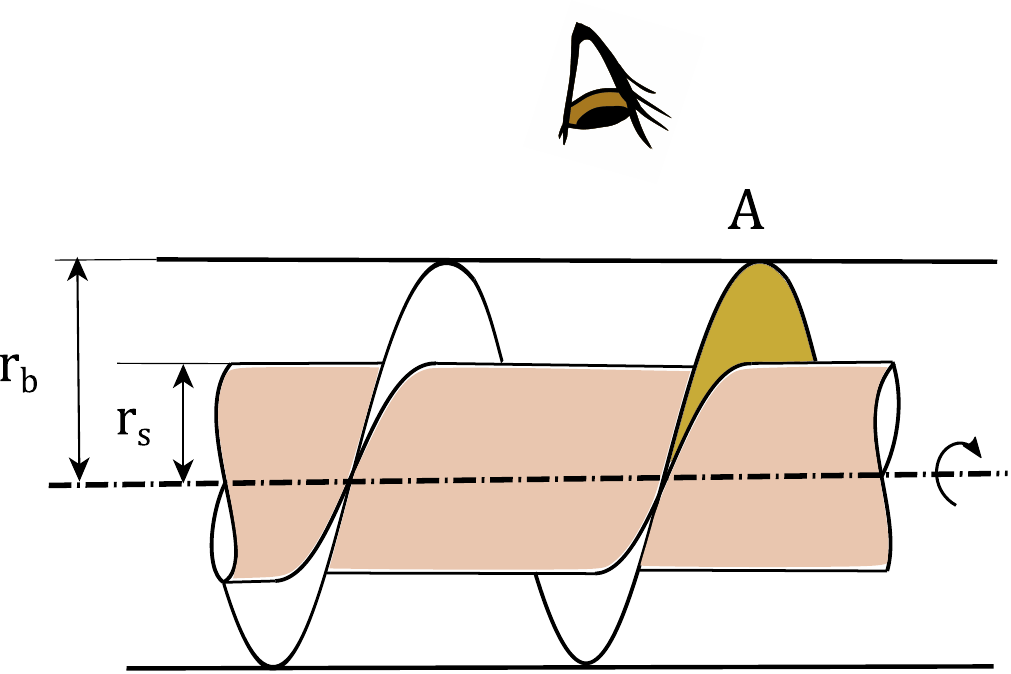}
			\caption{}
			\label{discharge_a}
		\end{subfigure}\\[1ex]
	\end{minipage}%
	\begin{minipage}{.5\linewidth}
		\centering
		\begin{subfigure}{\linewidth}
			\includegraphics[width=\linewidth]{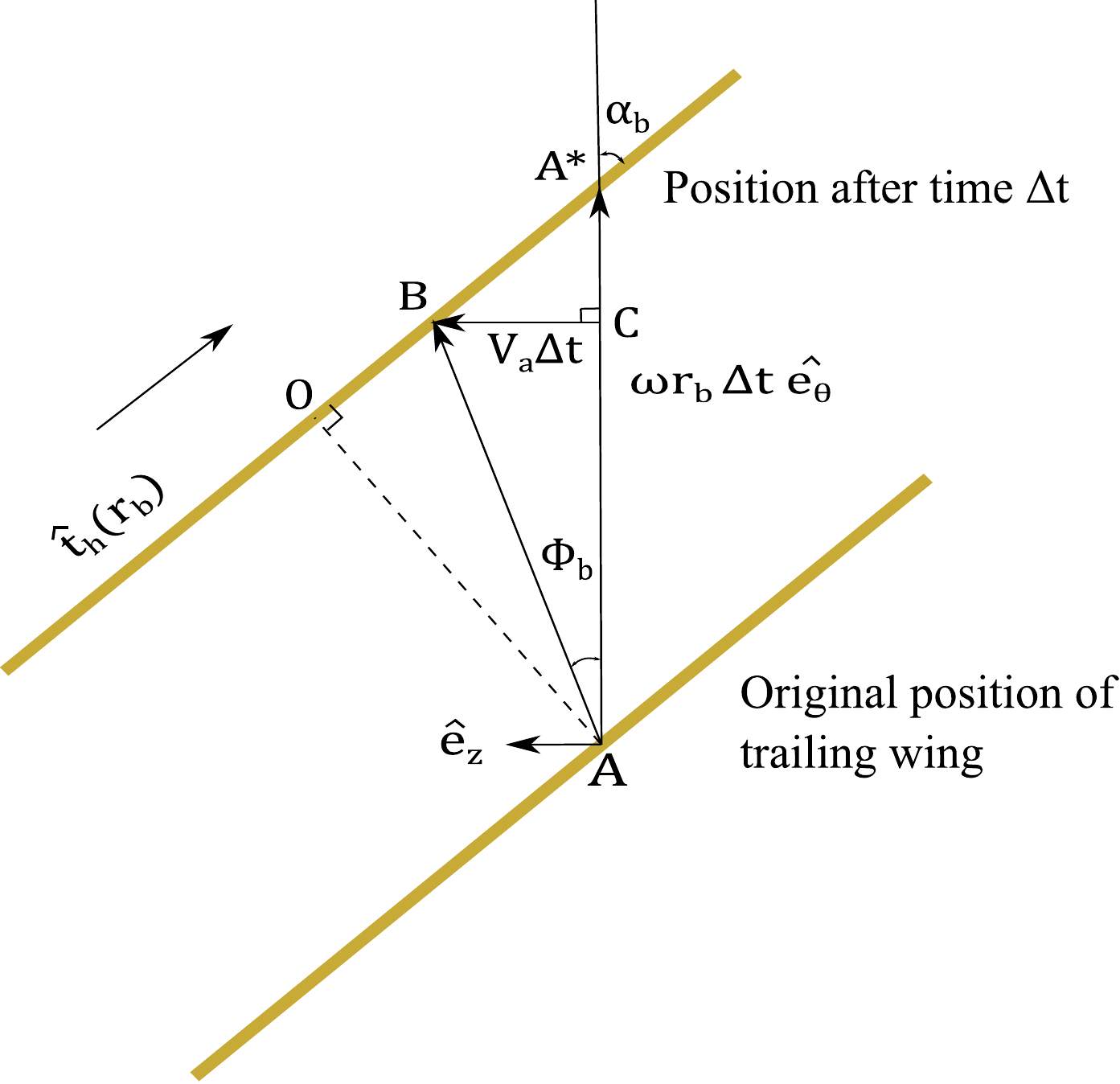}
			\caption{}
			\label{discharge_b}
		\end{subfigure}\\[1ex]
	\end{minipage}
	
	\caption{(a) The direction of viewing chosen to study the displacements of the trailing wing (highlighted in brown) and the solid plug adjoining it , (b) Displacements of the flight tip of the trailing wing at $A$, and the plug in a small time interval $\Delta t$}
	\label{discharge}
\end{figure}

We intend to relate the axial displacement of the element under consideration to the rotational speed of the screw. Consider Figure \ref{discharge}. The velocity triangle on the top is obtained by tracking the displacements of the tip of the screw flight at point $A$, and the solid plug adjoining it, in a small time interval of $\Delta t$. .\\

For a time interval, $\Delta t \to 0$, the trailing wing tip at $A$ moves along $\hat{e_\theta}$ by a distance \,$\omega r_b \Delta t$\, to land at $A^*$. The solid plug, as a consequence, moves from $A$ to $B$, in a direction that subtends an angle $\phi_b$ from $\hat{e_\theta}$. $\overrightarrow{CB}$ is the displacement that relates to the axial velocity in the calculation of discharge.

The discharge or the volumetric flux at any cross-section, after neglecting the blade thickness, is given by
\begin{equation}
Q = \nu V _ {a} \pi \left( r _ { b } ^ { 2 } - r _ { s } ^ { 2 } \right)
\end{equation}
where $\nu$ is the solids fraction of the packing inside the conveyor, and $V_a$ is the axial velocity of the plug.\\

The axial velocity can be further related to the azimuthal velocity of the tip of the screw flight as follows.
\begin{equation}
\frac { V _ {  a } } { V _ { b } } = \frac { BC } { A A ^ { * } } = \frac { BC } { A C + C A ^ { * } } = \frac { 1 } { \frac { 1 } { \tan \phi _ { b } } + \frac { 1 } { \tan \alpha _ { b } } } 
\end{equation}

\begin{equation}
V_{a} = V_{b} \frac { \tan \phi _ { b } \tan \alpha _ { b } } { \tan \alpha _ { b } + \tan \phi _ { b } }
\label{ax_vel_exp}
\end{equation}\\

Further, the azimuthal velocity of the plug at the barrel can be related to its axial velocity as follows.
\begin{equation}
V_{\theta} = \frac{V_{a}}{\tan \phi_b} =V_{b} \frac { \tan \alpha _ { b } } { \tan \alpha _ { b } + \tan \phi _ { b } }
\end{equation}\\

Substituting $V _ { b } = \omega r _ { b }$ in Eq. \ref{ax_vel_exp}, the expression for discharge attains the following form. 

\begin{equation}
{Q = \nu \omega  r _ { b } \, \frac { \tan \phi _ { b } \tan \alpha _ { b } } { \tan \alpha _ { b } + \tan \phi _ { b } } \, \pi \left( r _ { b } ^ { 2 } - r _ { s } ^ { 2 } \right)}
\label{discharge_exp}
\end{equation}

\vspace{10 pt}
The rest of the exercise that follows will be for determining $\phi_b$.\\

The magnitude and direction of forces acting on the white sub-elemental strips as shown in Figure \ref{element} are tabulated in Table \ref{table_forces}. The number in each subscript denotes the surface on which the force is acting. The numbers 1,2,3 and 4 refer to the surfaces of the element adjacent to the barrel, shaft, leading and trailing wing respectively; the letters $n$ and $t$ denote the normal and the tangential components of the force respectively.

\begin{table*}
\centering
\setlength{\tabcolsep}{6pt} 
\renewcommand{\arraystretch}{1.0} 
\begin{tabular}{ l|c }
	\hline
	\textbf{Force} & \textbf{Direction} \\ 
	\hline \hline\\
	$dF_{1n}=P_b(r_b d\theta d\xi)$ & $-\hat{e_r}$ \\
	$dF_{1f}=\mu_b P_b(r_b d\theta d\xi)$ & $-\cos\phi_b \, \hat{e_\theta}-\sin\phi_b \, \hat{e_z}  $\\
	\hline\\    
	$dF_{2n}=P_s(r_s d\theta d\xi)$ & $\hat{e_r}$\\
	$dF_{2f}=\mu_s P_s(r_s d\theta d\xi)$ & $\hat{t_h}(r_s)=\cos\alpha_s \, \hat{e_\theta}-\sin\alpha_s \,\hat{e_z}$\\
	\hline\\
	$dF_{3n}=P_{lw}\left(\left(\frac{rd\theta}{cos\alpha}\right)dr\right)$&$\hat{n_h}(r,\theta)=-\sin\alpha \, \hat{e_\theta}-\cos\alpha \, \hat{e_z}$\\
	$dF_{3f}=\mu_w P_{lw}\left(\left(\frac{rd\theta}{cos\alpha}\right)dr\right)$&$\hat{t_h}(r)=\cos\alpha \, \hat{e_\theta}-\sin\alpha \,\hat{e_z}$\\
	\hline\\
	$dF_{4n}=P_{tw}\left(\left(\frac{rd\theta}{cos\alpha}\right)dr\right)$&$-\hat{n_h}(r,\theta)=\sin\alpha \, \hat{e_\theta}+\cos\alpha \, \hat{e_z}$\\
	$dF_{4f}=\mu_w P_{tw}\left(\left(\frac{rd\theta}{cos\alpha}\right)dr\right)$&$\hat{t_h}(r)=\cos\alpha \, \hat{e_\theta}-\sin\alpha \,\hat{e_z}$\\
	\hline
\end{tabular}
	\caption{Various forces acting on the sub-elemental strips of the element under consideration, and their respective directions}
	\label{table_forces}
\end{table*}
 \vspace{10pt}
 
Note that it is only in the direction of frictional force acting on the plug at the barrel surface that the angle $\phi_b$ appears. After integrating the forces within suitable limits on all the surfaces of the element, the linear momentum balances along $\hat{e_r}$, $\hat{e_\theta}$, $\hat{e_z}$ result in the following equations.\\

\vspace{10pt}
\textbf{$\hat{e_r}$:}
\begin{equation}
{r_s\overline{P_s(z)}-r_b\overline{P_b(z)}+\overline{P}=0}
\label{lm1}
\end{equation}\\

\textbf{$\hat{e_\theta}$: }
\begin{equation}
{\begin{array}{c}
	{-\cos\phi_b \, \mu_b r_b \overline{P_b(z)}+\mu_s r_s \overline{P_s(z)}\cos\alpha_s-\frac{p}{2\pi} \overline{P_{lw}(r)}}\\\\
	{+\mu_w(\overline{rP_{lw}(r)})+\frac{p}{2\pi}\overline{P_{tw}(r)}+\mu_w (\overline{rP_{tw}(r)})=0}\\
	\end{array}}
\end{equation}\\

\textbf{$\hat{e_z}$: }
\begin{equation}
{\begin{array}{c}
	{-\mu_br_b\sin\phi_b \, \overline{P_b(\xi)}-\mu_sr_s \sin\alpha_s \overline{P_s(\xi)}-\overline{rP_{lw}(r)}}\\\\
	{-\frac{p\mu_w}{2\pi} \,\overline{P_{lw}(r)}+\overline{rP_{tw}(r)}-\frac{p\mu_w}{2\pi} \,\overline{P_{tw}(r)}=0}\\
	\end{array}}
\label{lm3}
\end{equation}\\

where the bar over any quantity $\phi(x)$ is it's integral with respect to $x$, given by $\overline{\phi(x)}=\int_{x_i}^{x_f} \phi(x) \, dx$. Thus $\overline{P(r,z)}$ is the zeroth moment of pressure on the rz plane. \\

Now, the contribution to the total torque on the element from the forces acting on the white sub-elemental strips are tabulated in Table \ref{table_torques}. The subscripts carry the same meaning as that mentioned earlier.\\

\begin{table}
	\centering
	\setlength{\tabcolsep}{6pt} 
	\renewcommand{\arraystretch}{0.8} 
\begin{tabular}{l|c}
	\hline
	\textbf{Torque} & \textbf{Position vector $\times$ Force} \\ 
	\hline \hline\\
	$dT_{1n}$&$dF_{1n}(z \, \hat{e_\theta})$\\
	$dT_{1f}$&$dF_{1f}(-z \cos\phi_b \, \hat{e_r}+r_b \sin\phi_b \, \hat{e_\theta}-r_b \cos\phi_b \, \hat{e_z})$\\
	\hline\\ 
	$dT_{2n}$&$dF_{2n}(-z \hat{e_\theta})$\\
	$dT_{2f}$&$dF_{2f} (z \cos\alpha_s \hat{e_r}+r_s \sin\alpha_s \hat{e_\theta}+r_s \cos\alpha_s \hat{e_z})$\\
	\hline\\ 
	$dT_{3n}$&$dF_{3n}(r\cos\alpha \hat{e_\theta}-r\sin\alpha \hat{e_z})$\\
	$dT_{3f}$&$dF_{3f}(r\cos\alpha \hat{e_z}+r\sin\alpha \hat{e_\theta})$\\
	\hline\\ 
	$dT_{4n}$&$dF_{4n}(p\sin\alpha \hat{e_r}-r\cos\alpha \hat{e_\theta}+r\sin\alpha \hat{e_z})$\\
	$dT_{4f}$&$dF_{4f}(p\cos\alpha \hat{e_r}+r\sin\alpha \hat{e_\theta}+r\cos\alpha \hat{e_z})$\\
	\hline
\end{tabular}
\caption{Torques arising from the forces acting on the sub-elemental strips of the element under consideration}
\label{table_torques}
\end{table}
After integrating the torques within suitable limits on all the surfaces of the element, the angular momentum balances along $\hat{e_r}$, $\hat{e_\theta}$, $\hat{e_z}$ would result in the following equations.\\

\vspace{10pt}
\textbf{$\hat{e_r}$:}
\begin{equation}
\begin{array}{c}
{-\cos\phi_b \mu_b r_b \overline{z P_b(z)}+\cos\alpha_s \mu_s r_s \overline{z P_s(z)} +\frac{p^2}{2\pi} \overline{P_{tw}(r)}}\\\\
{+p\mu_w \overline{rP_{tw}(r)}=0}
\end{array}
\label{am1}
\end{equation}\\

\textbf{$\hat{e_\theta}$: }
\begin{equation}
{\begin{array}{c}
	{r_b \overline{z P_b(z)}+r_b^2 \sin\phi_b \mu_b \overline{P_b(z)}-r_s \overline{z P_s (z)}}\\\\
	{+\sin\alpha_s \mu_s r_s^2 \overline{P_s(z)}+\overline{r^2 P_{lw}(r)}+\frac{p}{2\pi} \mu_w \overline{rP_{lw}(r)}}\\\\
	{-\overline{r^2P_{tw}(r)}+\frac{p}{2\pi} \mu_w \overline{rP_{tw}(r)}=0}
	\end{array}}
\end{equation}\\

\textbf{$\hat{e_z}$:} 
\begin{equation}
{\begin{array}{c}
	{-r_b^2 \cos\phi_b \mu_b \overline{P_b(z)}+\mu_s \cos\alpha_s r_s^2 \overline{P_s(z)}-\frac{p}{2\pi} \overline{rP_{lw}(r)}}\\\\
	{+\mu_w \overline{r^2P_{lw}(r)}+\frac{p}{2\pi} \overline{rP_{tw}(r)}+\mu_w \overline{r^2P_{tw}(r)}=0}
	\end{array}}
\label{am3}
\end{equation}\\

The linear momentum balances (Eq.\ref{lm1} - Eq.\ref{lm3}) and the angular momentum balances (Eq.\ref{am1} - Eq.\ref{am3}) together contain 12 unknowns including {${\phi_b}$} (the angle of interest), but we have only 6 equations available to us. However, the simplification that the friction coefficients on the flight and the shaft surfaces are zero while that on the barrel surface is still some finite value, allows us to determine $\phi_b$ (even though we still can not solve for all the unknowns involved). Hence, we first consider the case $\mu_s$ and $\mu_w = 0$.\\\\
The linear momentum balance along $\hat{e_z}$ reduces to
\begin{equation}
{
	-\mu_br_b\sin\phi_b \, \overline{P_b(z)}-\overline{rP_{lw}}(r)+\overline{rP_{tw}(r)}=0.}
\label{lin_mom_z}
\end{equation}\\
The angular momentum balance along $\hat{e_z}$ reduces to
\begin{equation}
{
	-r_b^2 \cos\phi_b \mu_b \overline{P_b(z)}-\frac{p}{2\pi} \overline{rP_{lw}(r)}+\frac{p}{2\pi} \overline{rP_{tw}(r)}=0.
}
\label{ang_mom_z}
\end{equation}\\
It is to be noted that in Table \ref{table_forces}, the axial component of forces from the flight surfaces have an additional explicit $r$ dependence as compared to the azimuthal component of these forces, which contribute to the angular momentum along $\hat{e_z}$. That is why, in Eq. \ref{ang_mom_z}, the order of pressure moments ends up being same as in Eq. \ref{lin_mom_z}. It is precisely the nature of the geometry that allows us to obtain the following relation.
\begin{equation}
{\cot\phi_b=\frac{p}{2\pi r_b}=\tan \alpha_b}
\label{key_relation}
\end{equation}\\
For a given $\alpha_b$, the angle $\phi_b$ obtained from the above equation is also the one that maximizes the discharge given by Eq. \ref{discharge_exp}. This is consistent with the expectation of the slip between the flights and the material, which is  responsible for the throughput in the axial direction, to be maximum in the absence of friction.
Using Eq. \ref{key_relation}, the expression for discharge  simplifies to
\begin{equation}
{
\begin{aligned}
Q &= \nu \,\omega  r _ { b } \, \sin\alpha_b \, \cos\alpha_b \, \pi \left( r _ { b } ^ { 2 } - r _ { s } ^ { 2 } \right)\\ 
	&=\nu \,\omega  r _ { b } \, \left( \frac{2\pi \left(p/r_b\right)}{\left(p/r_b\right)^2+(2\pi)^2} \right) \, \pi \left( r _ { b } ^ { 2 } - r _ { s } ^ { 2 } \right)\\
	&=\nu \,\omega  r_b \, f(p/d) \, \pi \left( r _ { b } ^ { 2 } - r _ { s } ^ { 2 } \right)
\end{aligned}}
\end{equation}\\
where,\\
\begin{equation}
f(p/d)=\left( \frac{4\pi \left(p/d\right)}{4\left(p/d)\right)^2+(2\pi)^2} \right)
\end{equation}\\

\begin{figure}
    	\centering
        \includegraphics[width=0.6\linewidth]{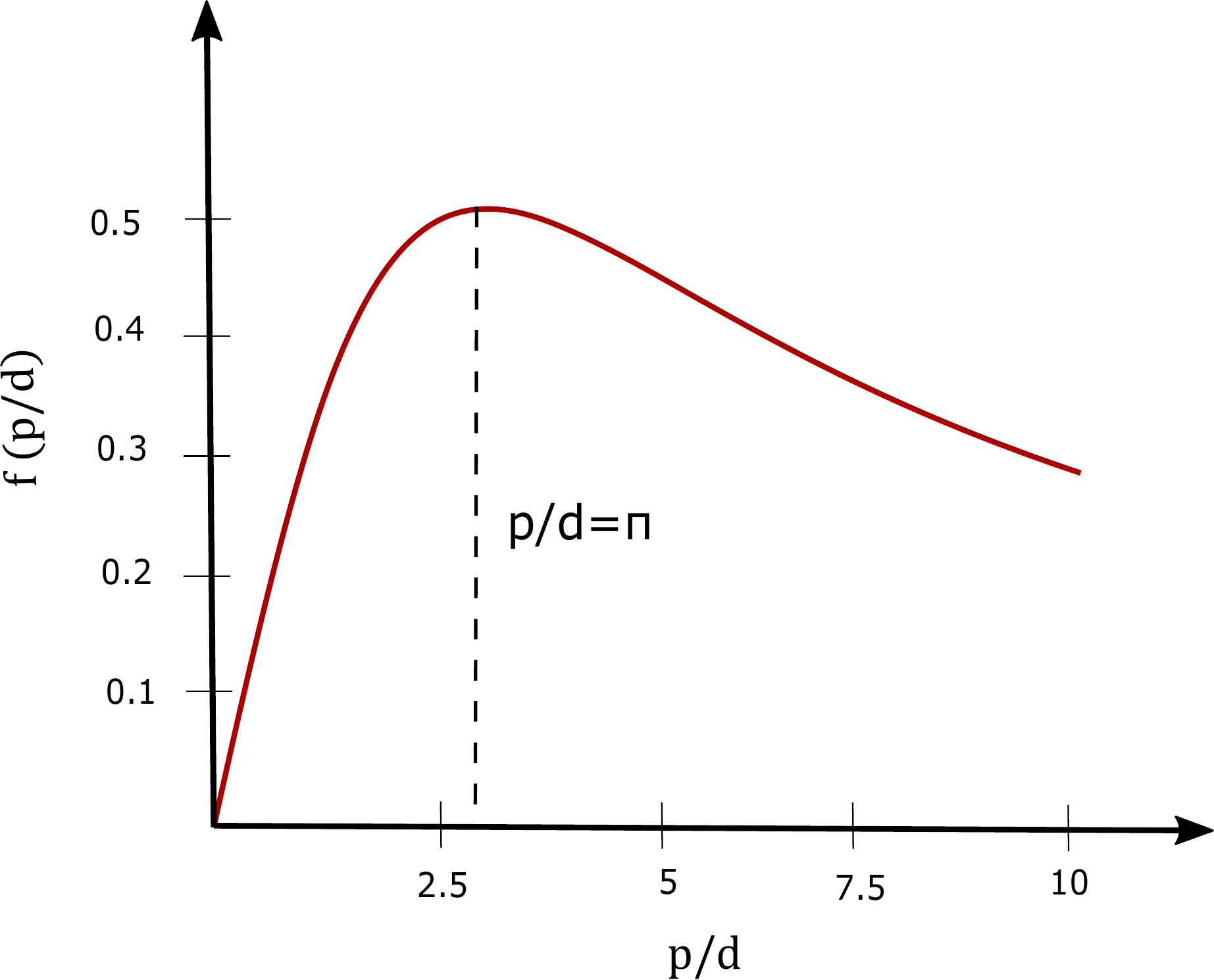}
		\caption{Variation of scaled discharge with the pitch to barrel diameter ratio }
		\label{plot_discharge}
\end{figure}

\noindent The discharge increases initially with the pitch to barrel diameter ratio, peaking at $\frac{p}{d}=\pi$ and then falls gradually with further increase in the ratio as shown in Figure \ref{plot_discharge}. The reason for the non-monotonic dependence will be discussed in \S 4.

\section{Results \& Discussions}
In this section, we present the results from our DEM simulations and interpret these results in the light of the understanding achieved from our frictionless screw model discussed in the preceding section. Even with the incorporation of friction on the screw surfaces in our gravity-less simulations, we will see that the kinematics of material inside is more or less plug-like. Shearing is only restricted to the regions very close to the screw and the barrel surfaces. Finally, the dependence of the average axial velocity on the pitch to barrel diameter ratio as obtained from simulations is compared with that from our model.
\subsection{Variation of quantities in the bulk (no gravity)}

\begin{figure}
		\begin{minipage}{.5\linewidth}
			\centering
		\begin{subfigure}{\linewidth}
			\includegraphics[width=0.9\linewidth]{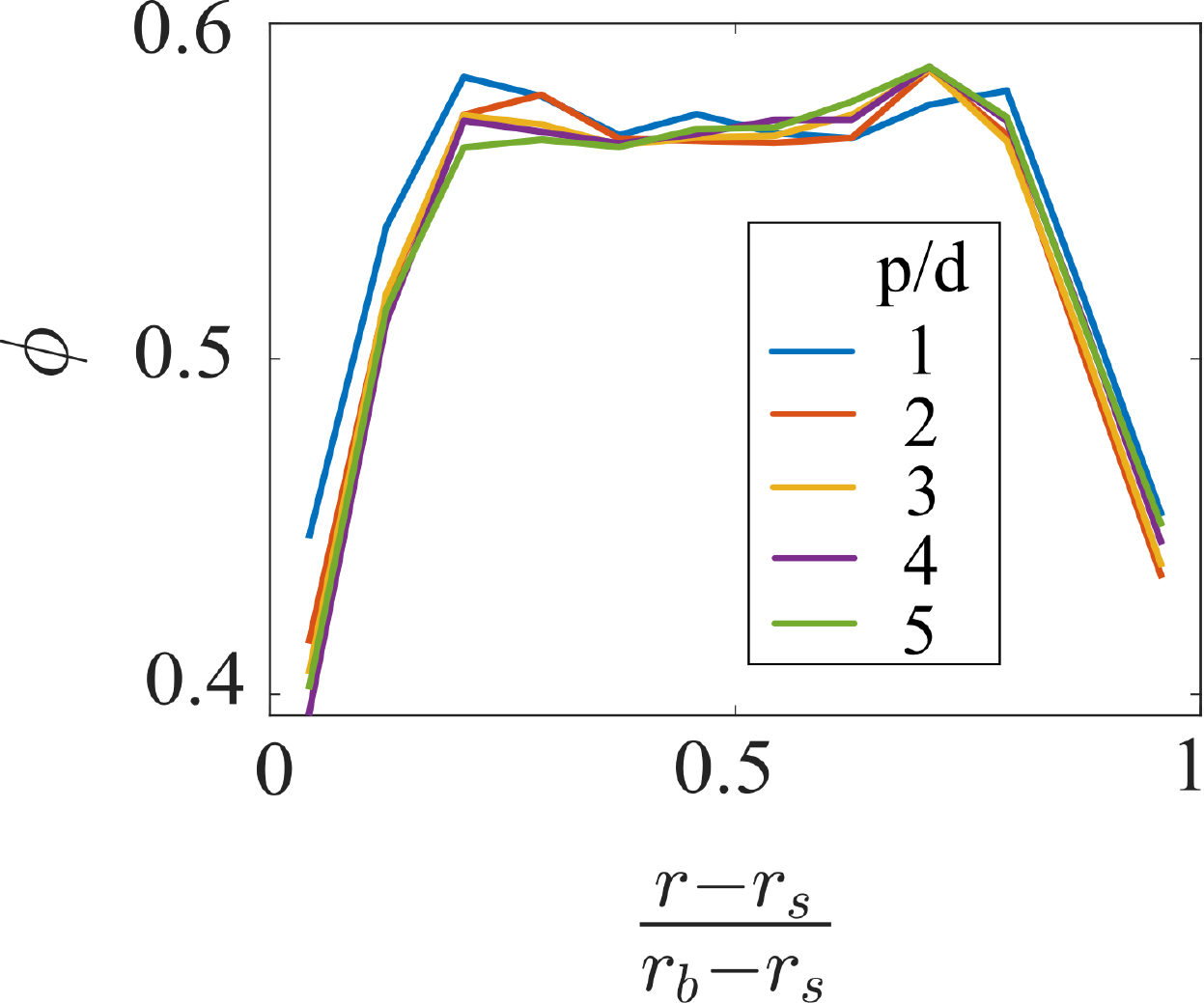}
			\caption{}
			\label{sf_1}
		\end{subfigure}\\[1ex]
		\begin{subfigure}{\linewidth}
			\includegraphics[width=0.90\linewidth]{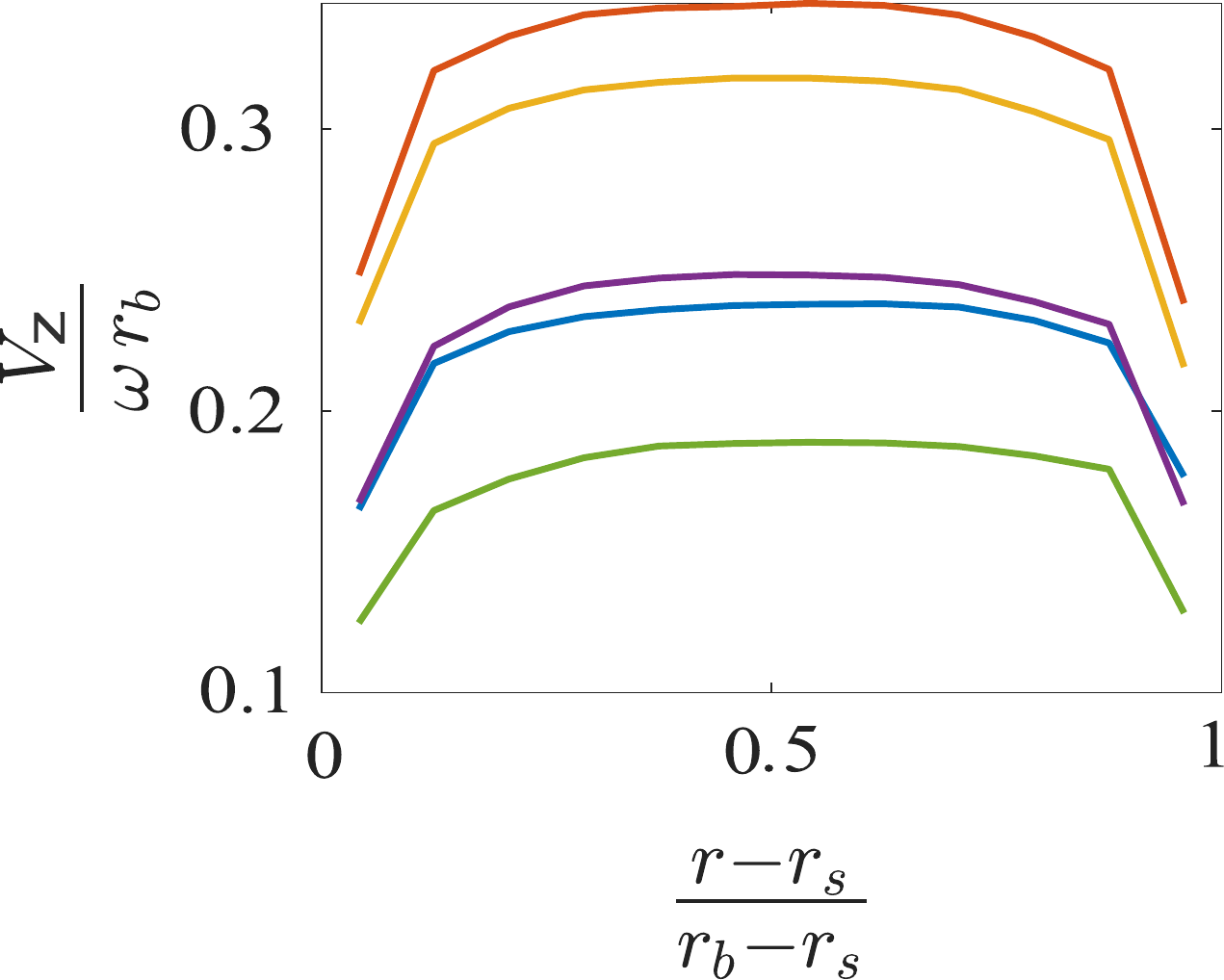}
			\caption{}
			\label{ax_vel_1}
		\end{subfigure}\\[1ex]
		\begin{subfigure}{\linewidth}
			\includegraphics[width=0.90\linewidth]{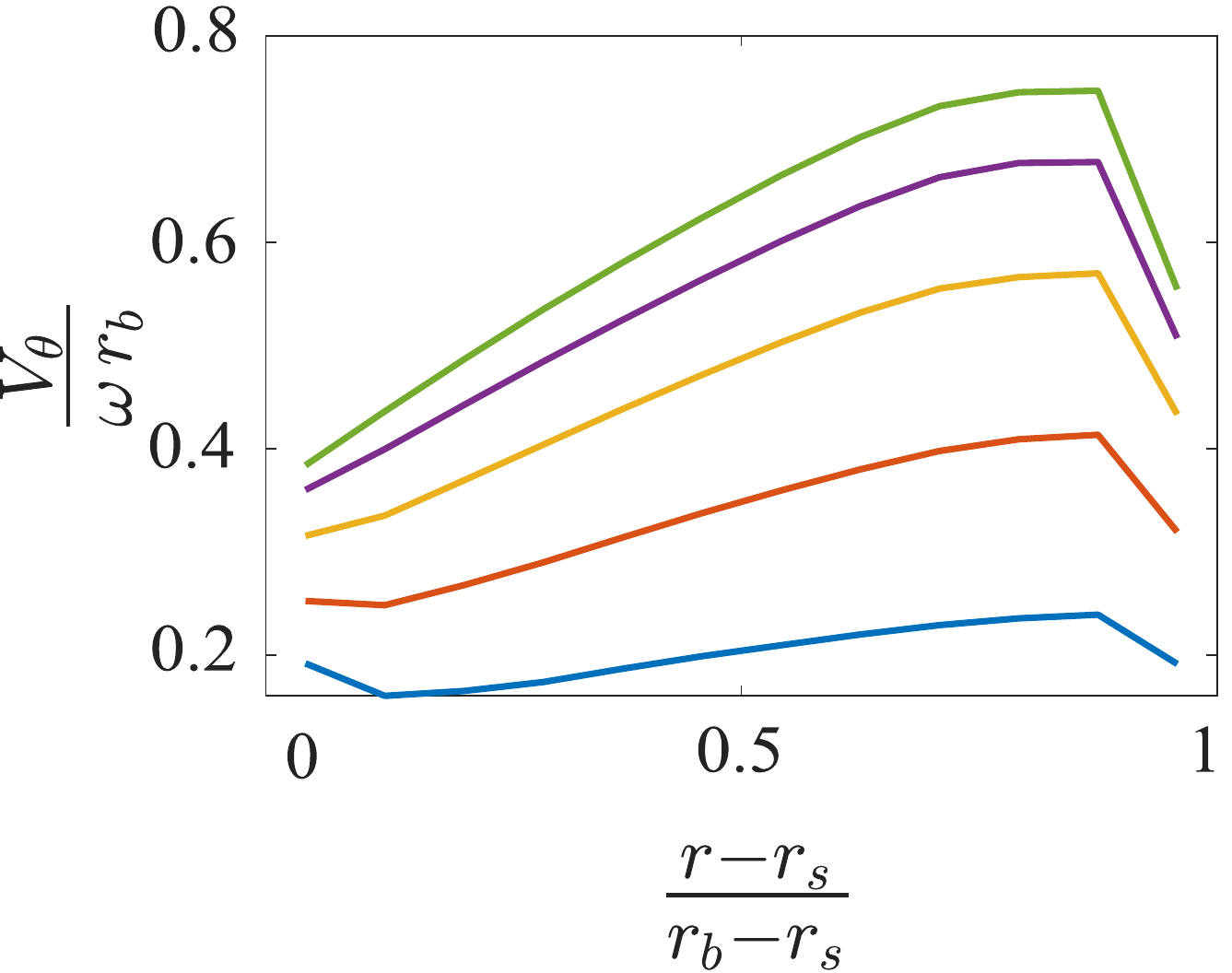}
			\caption{}
			\label{az_vel_1}
		\end{subfigure}	
	\end{minipage}%
	\begin{minipage}{.5\linewidth}
		\centering
	\begin{subfigure}{\linewidth}
		\includegraphics[width=0.90\linewidth]{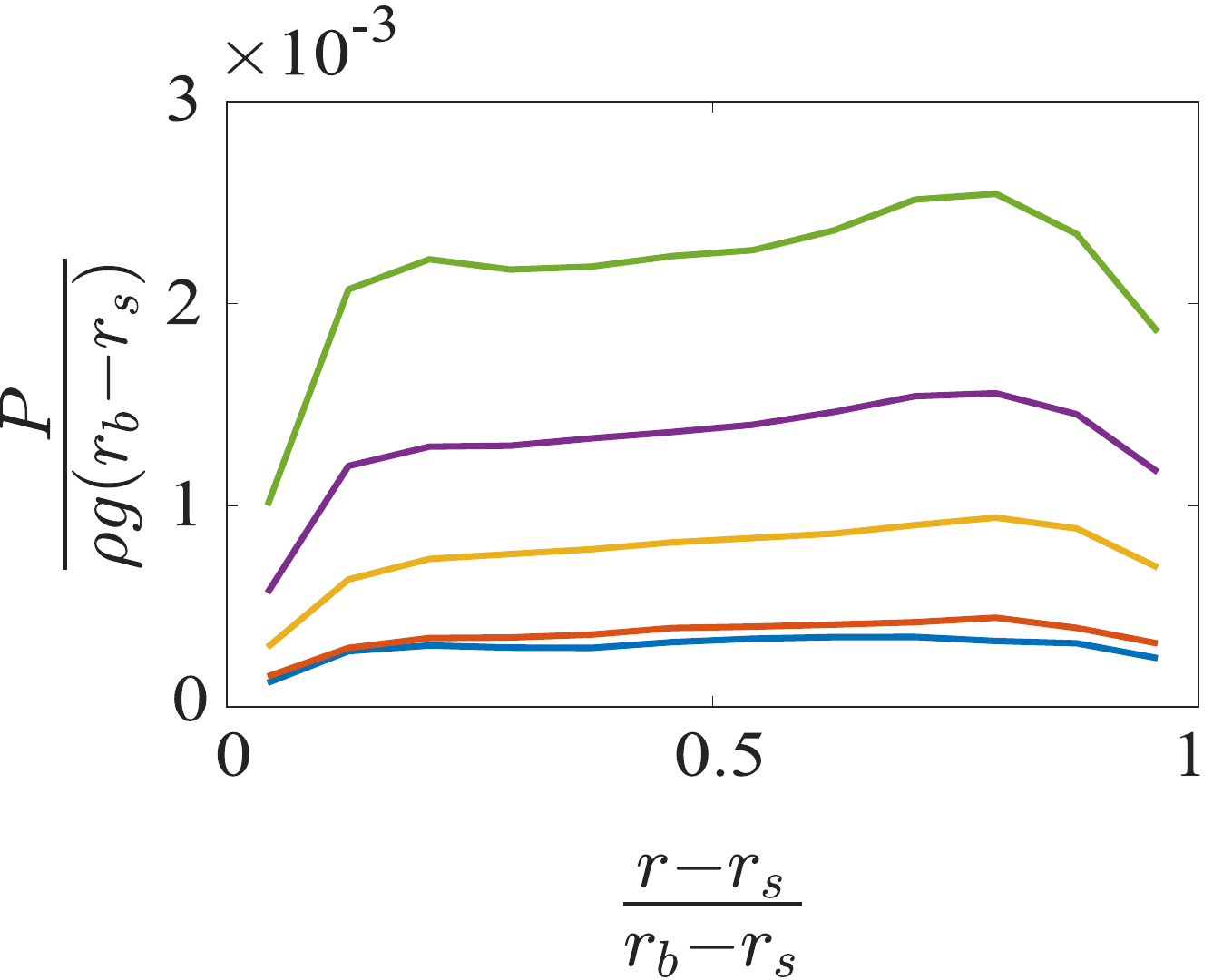}
		\caption{}
		\label{pressure}
	\end{subfigure}\\[1ex]
	\begin{subfigure}{\linewidth}
		\includegraphics[width=0.90\linewidth]{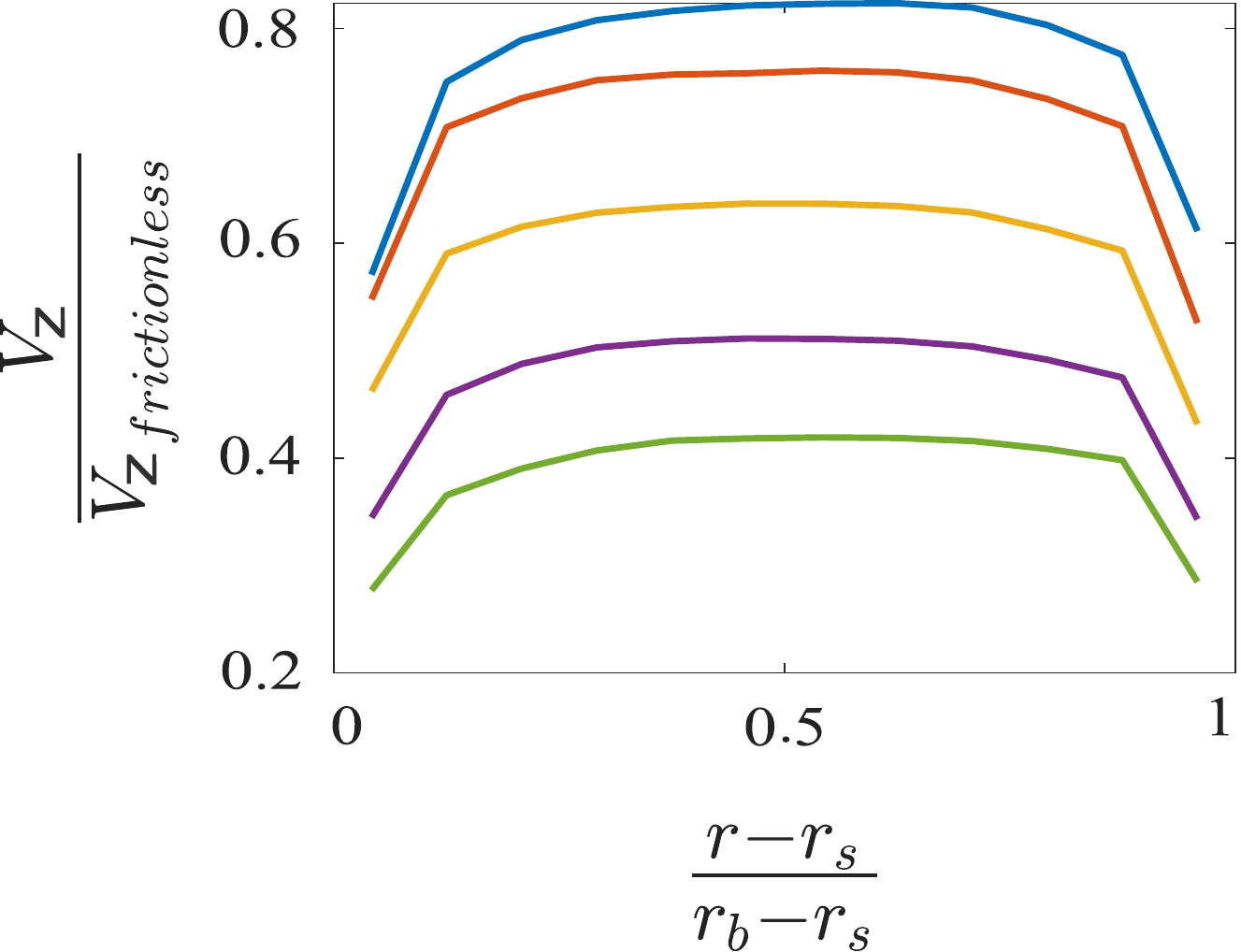}
		\caption{}
		\label{ax_vel_2}
	\end{subfigure}\\[1ex]
	\begin{subfigure}{\linewidth}
		\includegraphics[width=0.90\linewidth]{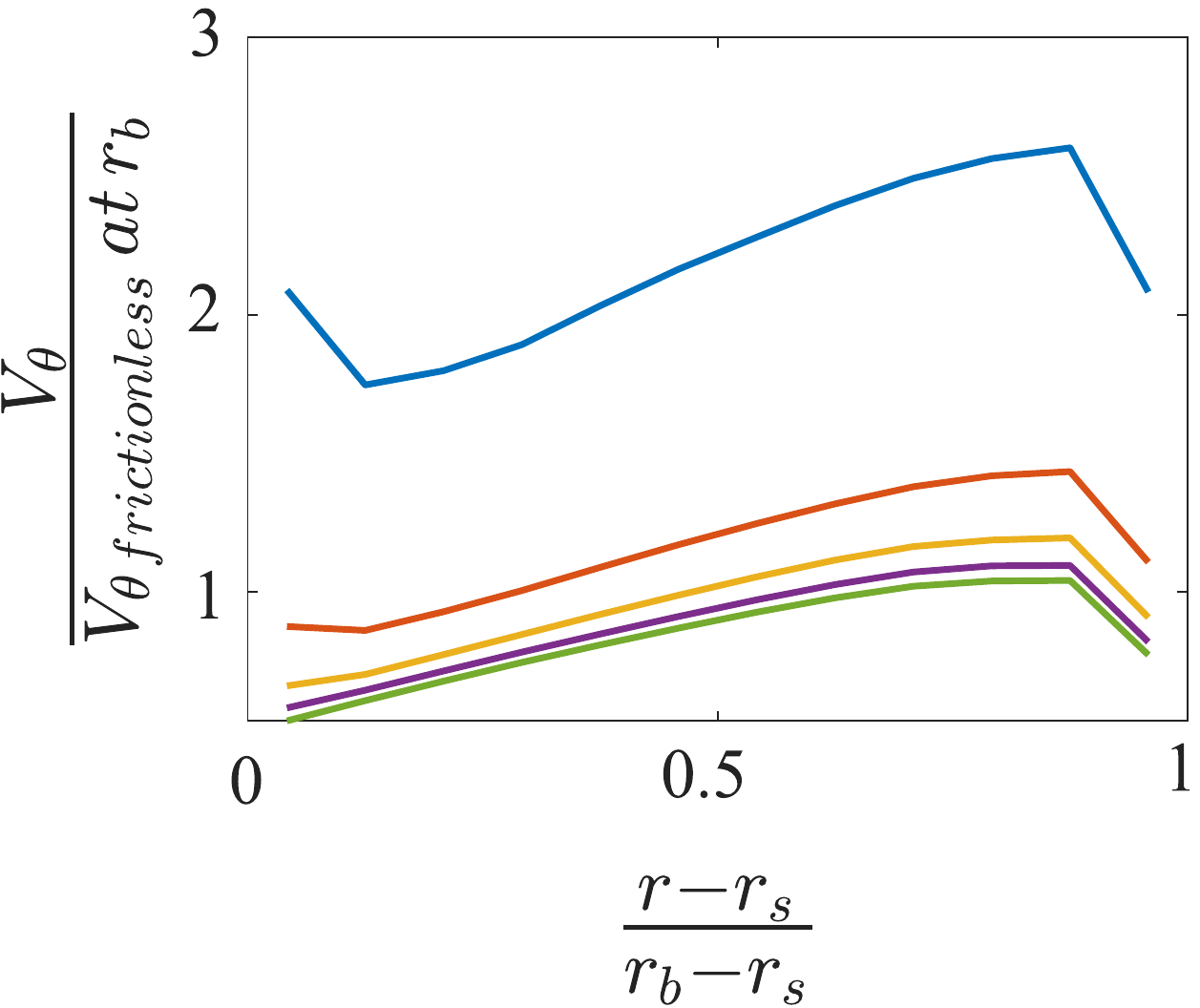}
		\caption{}
		\label{az_vel_2}
	\end{subfigure}
    \end{minipage}

	\caption{Radial variation of (a)solid fraction, (b)axial velocity scaled with the velocity of tip of the screw flight, (c)azimuthal velocity scaled with the velocity of tip of the screw flight, (d)pressure, (e)axial velocity scaled by the corresponding prediction from the frictional model for the same p/d ratio, (f)azimuthal velocity scaled by the azimuthal velocity at $r_b$ predicted by the frictional model for the same p/d ratio}
	\label{fig:test}
\end{figure}

Figure \ref{sf_1} shows that with changes in the pitch to diameter ratio of the screw, the radial variation of the solid fraction remains fairly unaffected. The solid fraction is relatively lower near the screw-shaft and the barrel, as is the case near a shearing 
surface \citep{bridgwater1980width,nedderman1980thickness,muhlhaus1987thickness,drake1990structural,oda1998microstructure}. Stress transmission in granular media takes place through force chains along certain preferential directions. Despite the presence of large normal stress differences, the pressure defined as one-third of the trace of the stress tensor, is still a useful quantity to examine. Figure \ref{pressure} clearly shows pressure being more or less constant in the core for each p/d ratio. However, the increase in pressure with the increase in p/d ratio is because of the consequent increase in the normal force imparted by the screw-flight to the grains, as a result of the increasing inclination of the flight.\\

The axial velocity for any p/d ratio of the screw is observed to be almost constant in the core. However, there are small regions near the screw-shaft and the barrel surfaces, where there are sharp gradients in the velocity. The variation of velocity in the core with p/d ratio is clearly non-monotonic as shown in Figure \ref{ax_vel_1}. Figure \ref{ax_vel_2} shows the deviation of the axial velocity for each p/d ratio from the results of the frictionless screw predicted by our model in \S3. With the incorporation of friction on the screw surfaces, the velocity drops for all the p/d ratios, but the drop becomes more and more pronounced as we move towards the higher ratios. \\

The azimuthal velocity for any p/d ratio of the screw increases linearly with the radial distance from the screw-shaft in the core, as shown in Figure \ref{az_vel_1}. The sudden drop in  velocity close to the barrel can be attributed to the friction resisting the relative motion at the barrel surface. Figure \ref{az_vel_2} shows the deviation of the azimuthal velocity for each p/d ratio from its frictionless counterpart predicted by our model. The azimuthal velocity adjoining the barrel surface for all p/d ratios greater than or equal to 2, is close to unity. However, for the smallest pitch to diameter ratio (p/d=1), the deviation is quite significant. This is because friction on the screw surface retards the sliding of material present inside, effectively causing the effect of screw rotation to be dominant in the azimuthal velocity.\\

The superposition of the axial and the azimuthal velocity profiles indicates that material points experience little or no deformation inside the conveyor, except very close to the shearing surfaces. This result supports the assumption of a solid-plug made in our model.\\


\subsection{Variation of quantities in the bulk (gravity present)}
\begin{figure}
	\begin{minipage}{.5\linewidth}
		\begin{subfigure}{\linewidth}
			\centering
			\includegraphics[width=.9\linewidth]{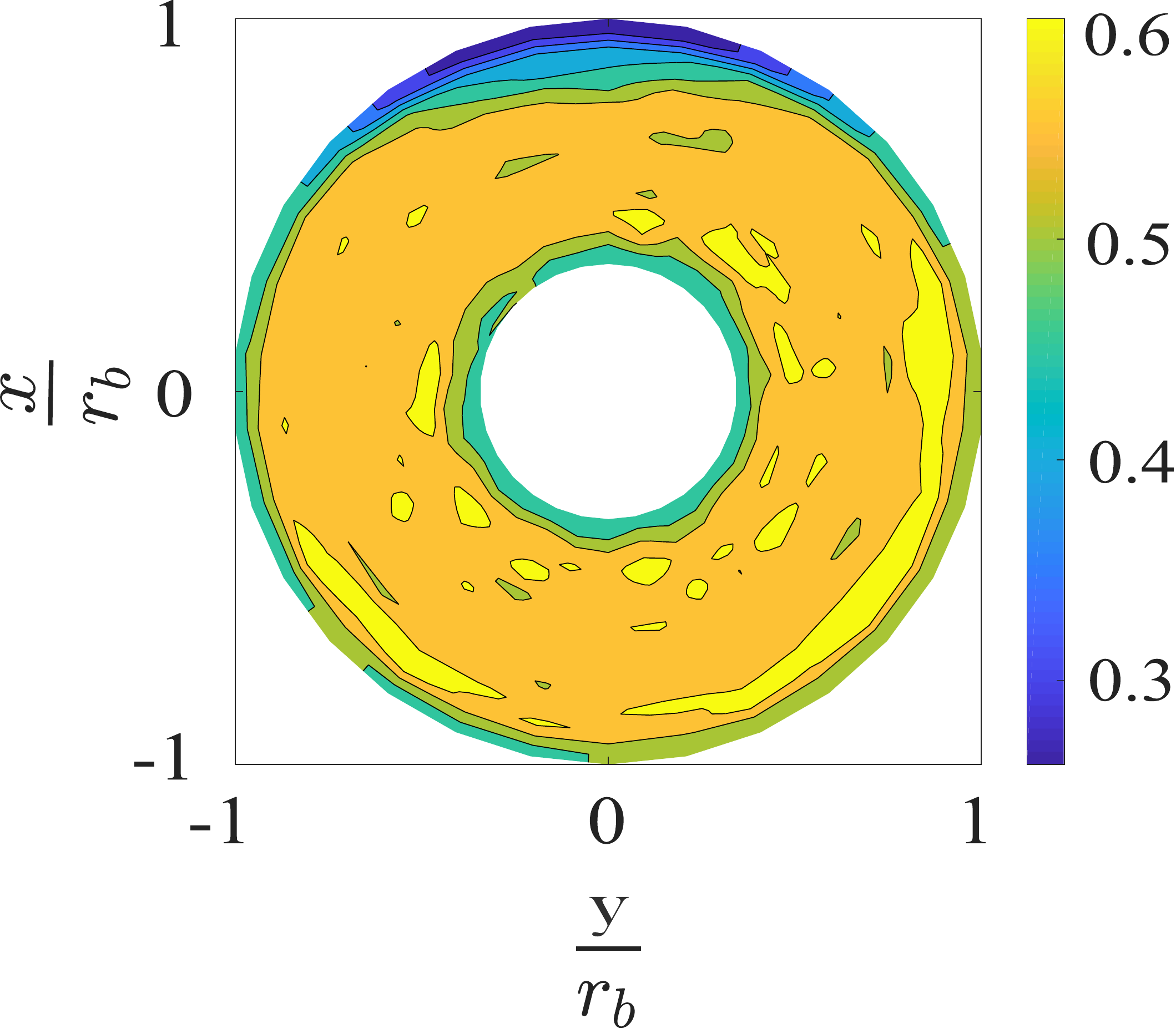}
			\caption{}
			\label{solid_fraction_g}
		\end{subfigure}\\[1ex]
		\begin{subfigure}{\linewidth}
			\centering
			\includegraphics[width=.9\linewidth]{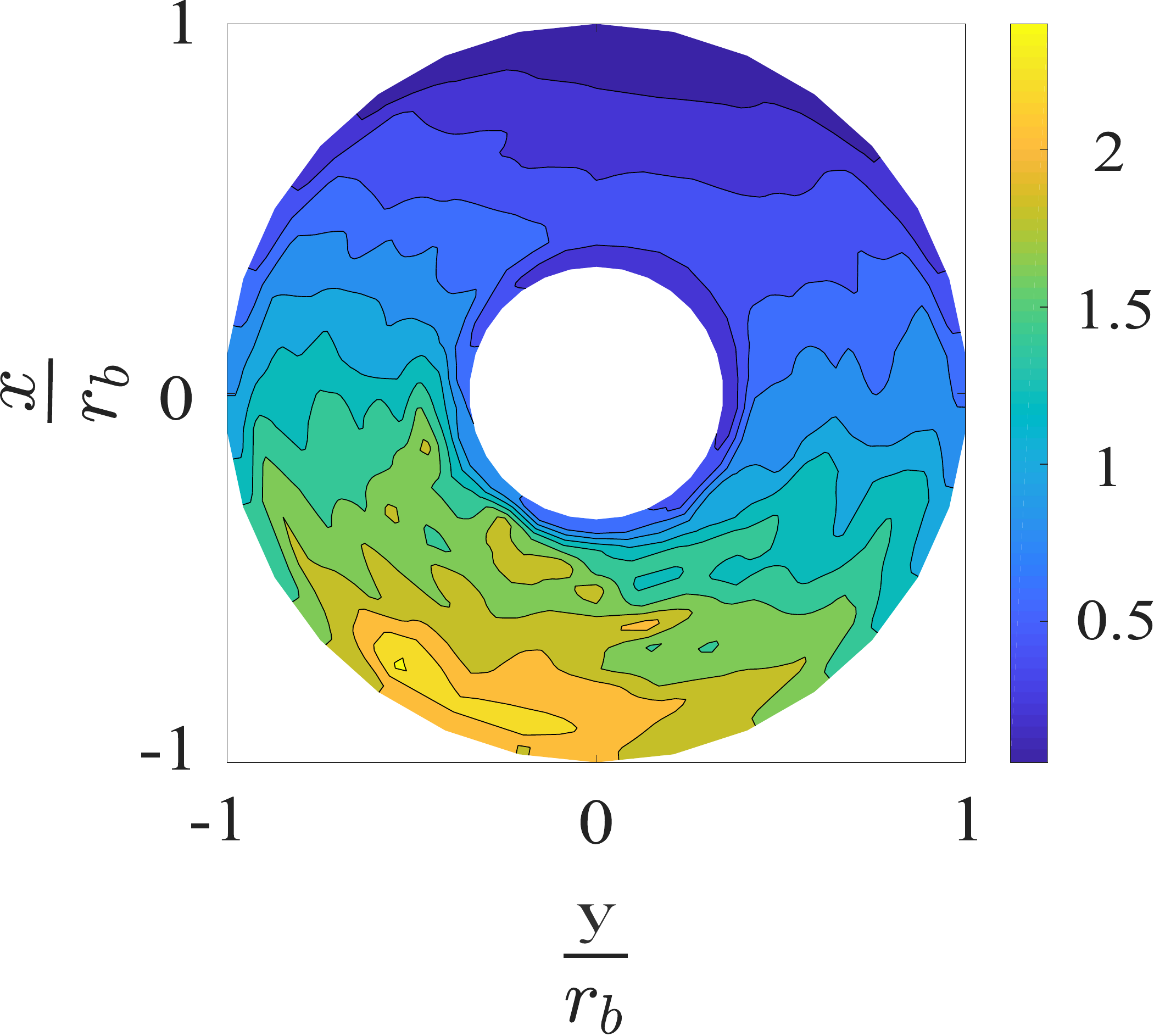}
			\caption{}
			\label{pressure_g}
		\end{subfigure}
	\end{minipage}%
	\begin{minipage}{.5\linewidth}
		\begin{subfigure}{\linewidth}
			\centering
			\includegraphics[width=.9\linewidth]{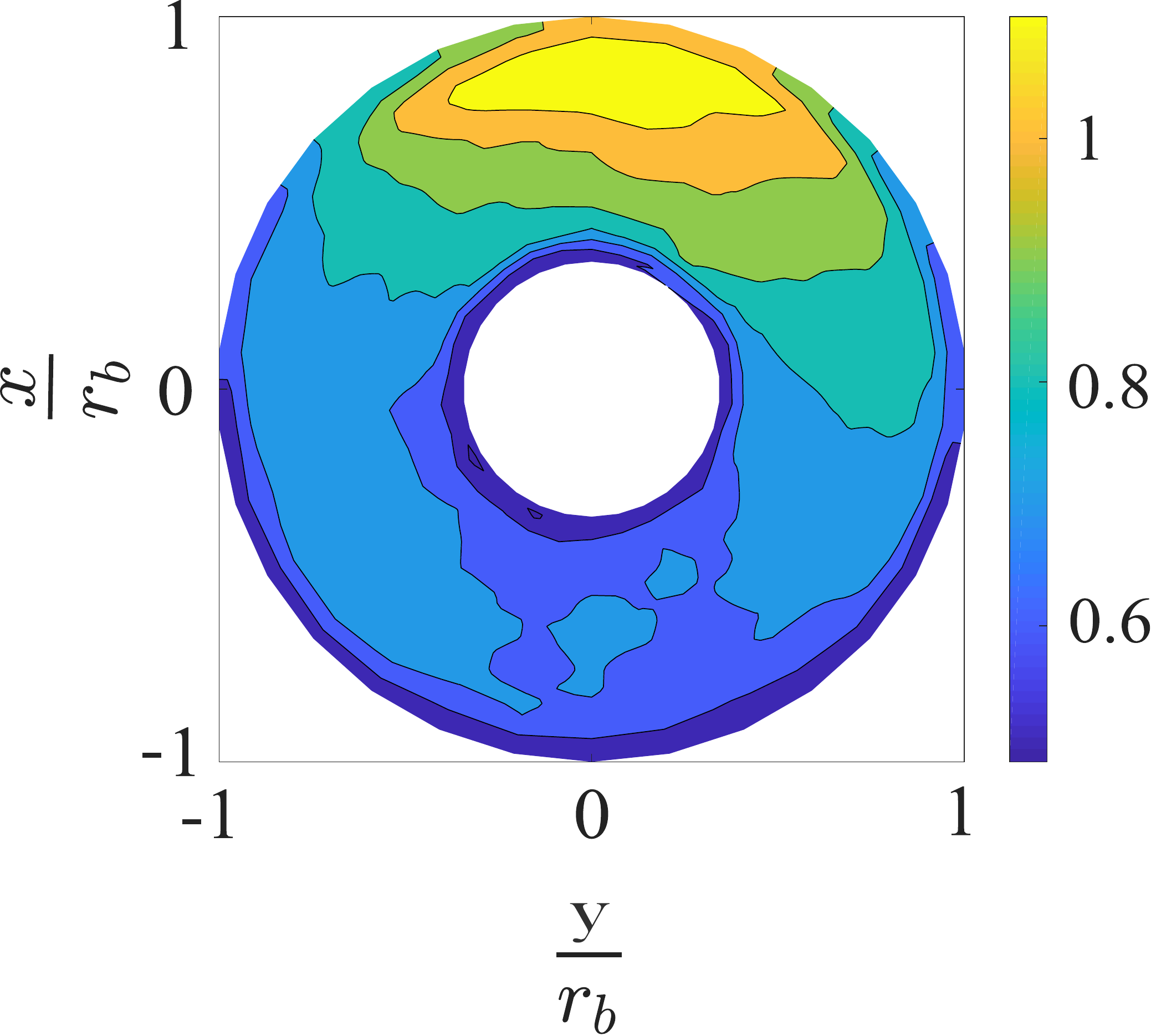}
			\caption{}
			\label{ax_vel_g}
		\end{subfigure}\\[1ex]
		\begin{subfigure}{\linewidth}
			\centering
			\includegraphics[width=.9\linewidth]{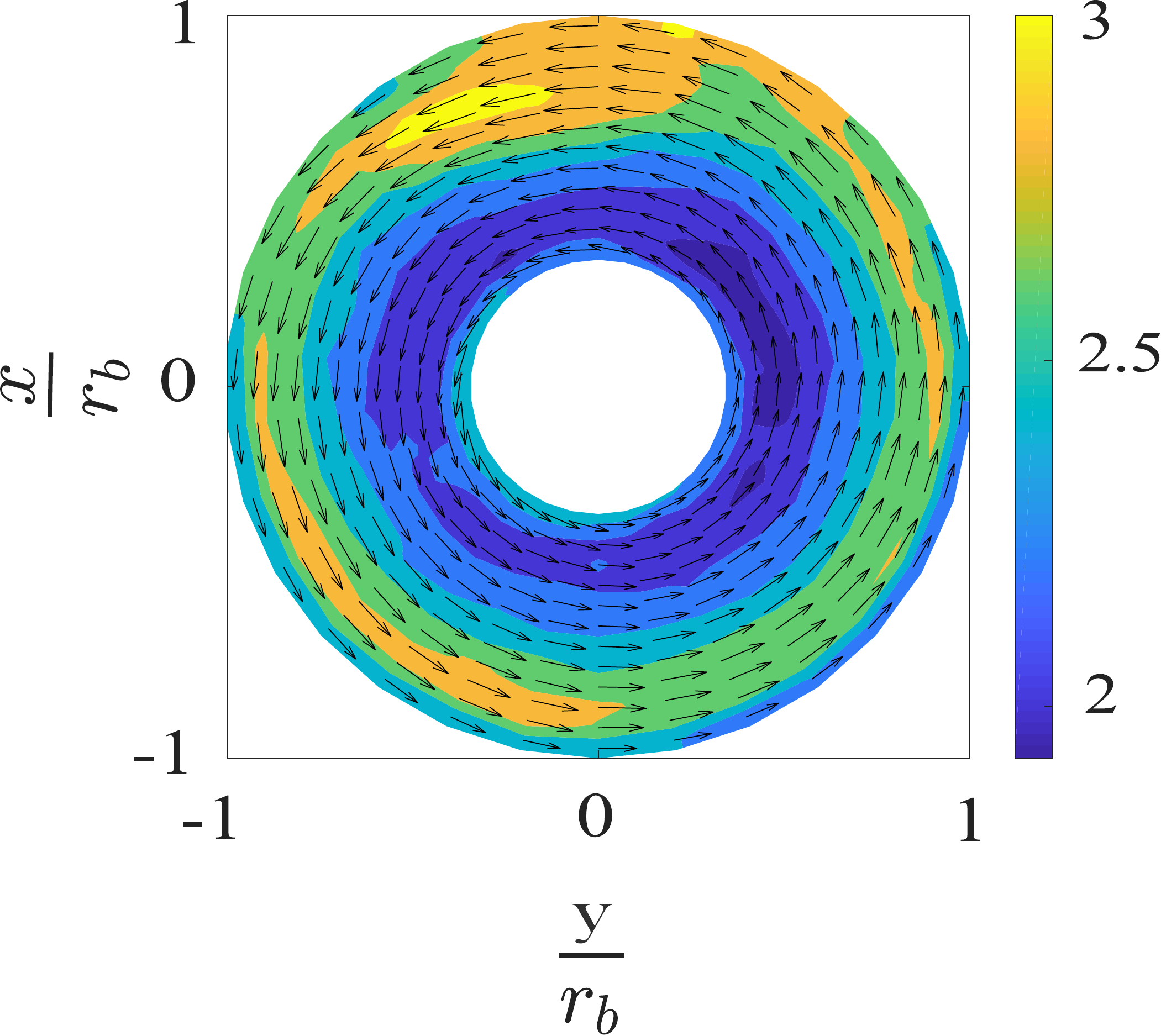}
			\caption{}
			\label{az_vel_g}
		\end{subfigure}
	\end{minipage}

	\caption{Variation of (a) solid fraction ($\phi$), (b) scaled pressure ($P/(\rho g (r_b-r_s))$), (c) scaled axial velocity ($V_z/V_{z \, model}$), and (d) scaled azimuthal velocity($V_\theta/V_{\theta \, model\, at \, r_b}$) in the region between the shaft and the barrel}
	\label{fig:test}
\end{figure}

Figure \ref{solid_fraction_g} shows that solid fraction is fairly constant in the bulk of the material, and lower near the shearing regions. Due to the compaction caused by gravity, the least solid fraction occurs near the barrel at the top. Figure \ref{pressure_g} shows the variation of pressure scaled with the hydrostatic pressure corresponding to a depth of $(r_b-r_s)$. The pressure is seen to rise with depth. However the gradient of pressure does not appear to be aligned with the direction of gravity. This could be a result of the directional bias introduced in the system because of rotation of the shaft. In Figure \ref{ax_vel_g}, the contour plot represents the variation of axial velocity scaled by its counterpart from our model in the limit of vanishing friction on the screw surfaces. The velocity is highest at the top where the solid fraction is low, and it decreases with depth from the barrel surface. Figure \ref{az_vel_g} is quiver plot of azimuthal  velocity scaled by its counterpart from our model in the limit of vanishing friction on the screw surfaces. The corresponding streamlines will be circular. The azimuthal velocity is lowest adjacent to the screw shaft and increases as we move radially outwards.

\section{Comparison of DEM result with theory} 
 \begin{figure}
	\centering
	\includegraphics[width=0.7\linewidth]{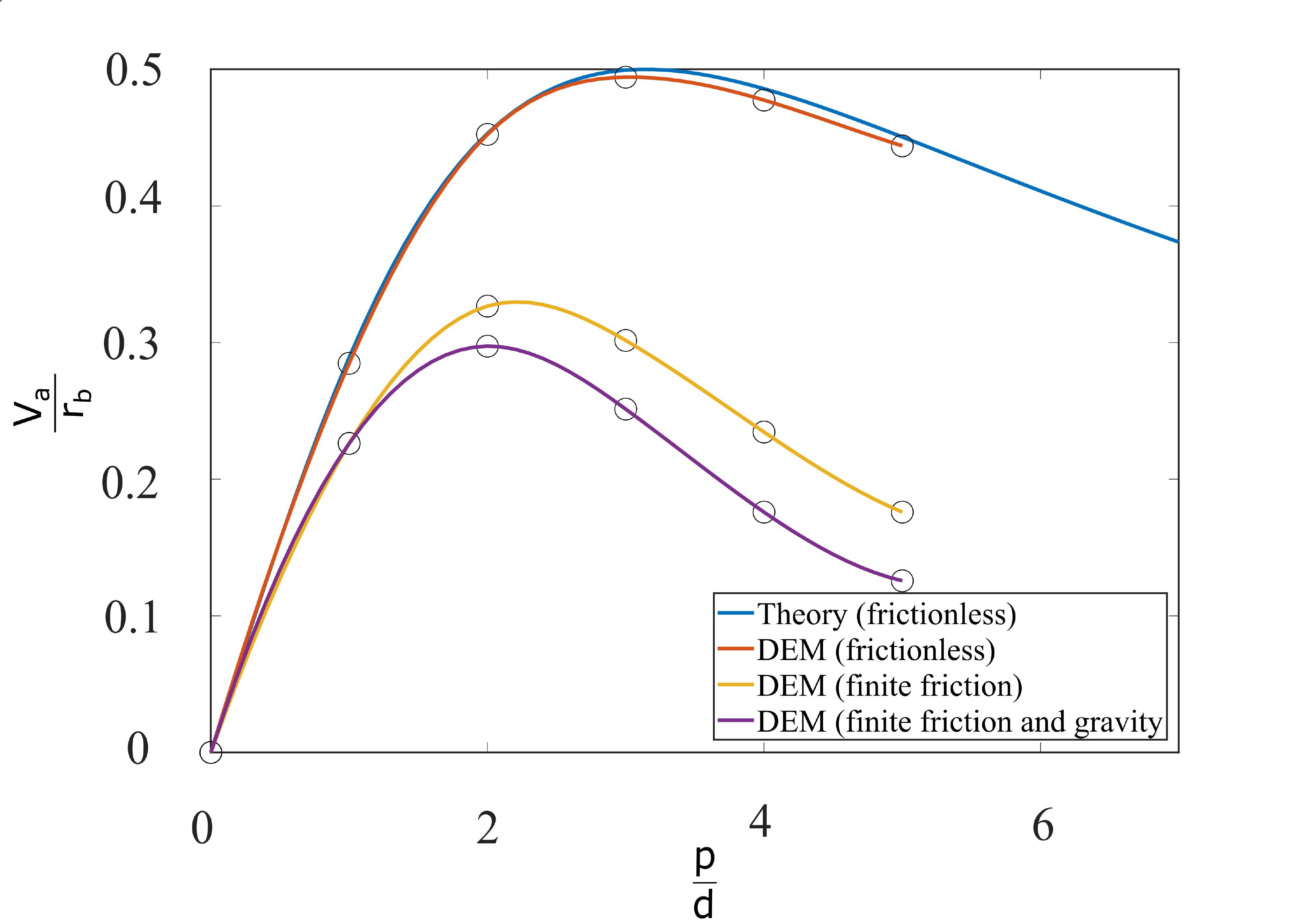}
	\caption{Variation of the average axial velocity of grains with the pitch to barrel diameter ratio. The circles represent results from the DEM simulations at specific p/d ratios. A cubic spline has been used for the purpose of interpolation.}
	\label{DEM_theory}
\end{figure}

Our model is able to capture the qualitative variation of the axial velocity with the pitch to barrel diameter ratio as shown in Figure \ref{DEM_theory}. The dependence of axial velocity of material inside a screw conveyor on its p/d ratio  can be understood by compounding the effects of two factors. The first factor is the magnitude of slip between the flight surface and the material, and the second factor is the component of that slip in the axial direction. For a given angular velocity of the screw shaft, as the flight becomes more and more aligned in the axial direction as a consequence of increasing p/d ratio, the magnitude of slip decreases where as the component of that slip in the axial direction increases. It is thus the interplay of these factors that results in a non-monotonic dependence of axial velocity on the p/d ratio. The slip goes from a maximum at p/d $\to$ 0 to a minimum at p/d $\to$ $\infty$. But when p/d $\to$ 0, as flight surface is virtually perpendicular to the axis of the screw, the slip does not have a component in the axial direction. Similarly, at the other extreme of p/d $\to$ $\infty$, the slip itself is zero and the net result is material simply going around in circles as if it were stuck between the flights. There is a sweet spot at p/d $=$ $\pi$ where the product of the slip velocity and the factor governing its component in the axial direction is maximized.\\

Furthermore, the presence of friction on the screw surfaces reduces the slip velocity which causes a drop in the axial velocity for all p/d ratios, as can be observed in Figure \ref{DEM_theory}. The figure also shows that the presence of gravity does not alter the dependence of axial velocity on the p/d ratio.

\section{Conclusion}
In the current work, we have developed a simple mechanical model based on rotation induced sliding to relate the axial velocity of material in a fully filled screw conveyor to its geometry. In our analysis, we did not assume any simplification of the geometry. We also allowed for the pressure variation on various surfaces of the screw. We then obtained a set of equations from the linear and angular momentum balances on a suitably chosen continuum element. Without knowledge of the radial variation of the pressure on the flight surface and the axial variation on the screw shaft and barrel, the number of unknowns were greater than the number of variables in the set of governing equations, but we could solve them in the limit of vanishing friction on the screw. An expression for discharge was obtained, suggesting a non-monotonic dependence on the pitch to barrel diameter ratio of the screw. A physical understanding of this dependence was also provided.\\

We then employed Discrete Element Method to understand the effect of friction on the screw surfaces, on the flow through the conveyor. The shearing was found to be confined near the walls , and the material in the core was found to conform with our assumption of a solid plug in the model. Simulations in the presence of gravity reveal that the flow is not exactly plug-like. However, the dependence of the average axial velocity on the geometry is similar to the case without gravity. The qualitative behavior of the dependence of average axial velocity on the pitch to barrel diameter ratio is same as the one predicted by our model for a frictionless screw.\\

 Now that we understand the dependence of flow solely on the geometry, we can also separate out the distortion caused to the flow by imposition of gravity. The scope of the current study was limited to a fully filled screw-conveyor. The flow profiles in a partially filled system will be more intricate. Nevertheless, we suspect that the dependence of average axial velocity on the geometry might not change significantly. However, this needs to be established via simulations or experiments.

\bibliographystyle{jfm}
\bibliography{references}

\appendix
\vspace*{30 pt}
\begin{center}
	\textbf{Appendix}
\end{center}
Here, we first describe the general framework of the DEM simulation technique and then the specific contact model chosen for our simulations. We provide the details about the dimensions of the geometry and also explain how the triangulation of the screw is achieved. We then provide an overview of the different stages in the simulation of our screw conveyor. Finally, we explain how the particle level information obtained from the simulations is translated into meaningful quantities such as average velocities, and stresses on the screw. 
\subsection{The Discrete Element Method}
The Discrete Element Method is a simulation technique in which the dynamics of a large ensemble of particles is studied by updating certain attributes of each individual particle with time. A modified version of the Verlet algorithm \citep{verlet1967computer} known as
velocity-Verlet \citep{swope1982computer}
is used to get the updated position, velocity, and force of each particle.

The contact forces during the collisions can be modeled by contact mechanics, such as a linear spring-dashpot-slider model. More complex models such as a nonlinear Hertz-dashpot-slider model, hysteretic spring model, and Hertz-Mindlin-Deresiewicz model can also be employed. \citep{walton1986viscosity,johnson1987contact,zhang1996calculation,shafer1996force,luding1998collisions,di2004analytical,di2004comparison}. The normal $\mathbf{F}_{ij}{^n}$ and tangential $\mathbf{F}_{ij}{^t}$ components of the contact forces can be modeled as the sum of spring forces ($\mathbf{F}_{ij}{^n}_{sp}$ and $\mathbf{F}_{ij}{^t}_{sp}$), which are conservative (elastic), and dashpot forces ($\mathbf{F}_{ij}{^n}_{da}$ and $\mathbf{F}_{ij}{^t}_{da}$), which are dissipative (viscous) by nature.
\begin{equation}
\mathbf{F}_{ij}{^n} = \mathbf{F}_{ij}{^n}_{sp} + \mathbf{F}_{ij}{^n}_{da}
\end{equation}
\begin{equation}
\mathbf{F}_{ij}{^t} = \mathbf{F}_{ij}{^t}_{sp} + \mathbf{F}_{ij}{^t}_{da}
\end{equation}
\\
In the linear spring-dashpot-slider model, the spring forces, $\mathbf{F}_{ij}{^n}_{sp}$ and $\mathbf{F}_{ij}{^t}_{sp}$ are linearly proportional to the extent of overlap according to Hooke's law. The dashpot forces, $\mathbf{F}_{ij}{^n}_{da}$ and $\mathbf{F}_{ij}{^t}_{da}$ are linearly proportional to the normal and tangential components of the relative velocity of particle j with respect to particle i at the contact point $\mathbf{v}_{ji}$. 

In the normal direction the interaction force is 
\begin{equation}
\mathbf{F}_{ij}{^n} = -k_n\delta_n\mathbf{n} - \xi_n \mathbf{v}_n
\end{equation}

where $k_n$ is the normal spring stiffness coefficient, $\delta_n$ is the deformation in the normal direction, and $\xi_n$ is the normal damping coefficient.
\begin{equation}
\delta_n=(R_i+R_j)-\left\vert{r_j-r_i}\right\vert 
\end{equation}

\begin{figure}
	\begin{center}
		\includegraphics[width=0.4\linewidth]{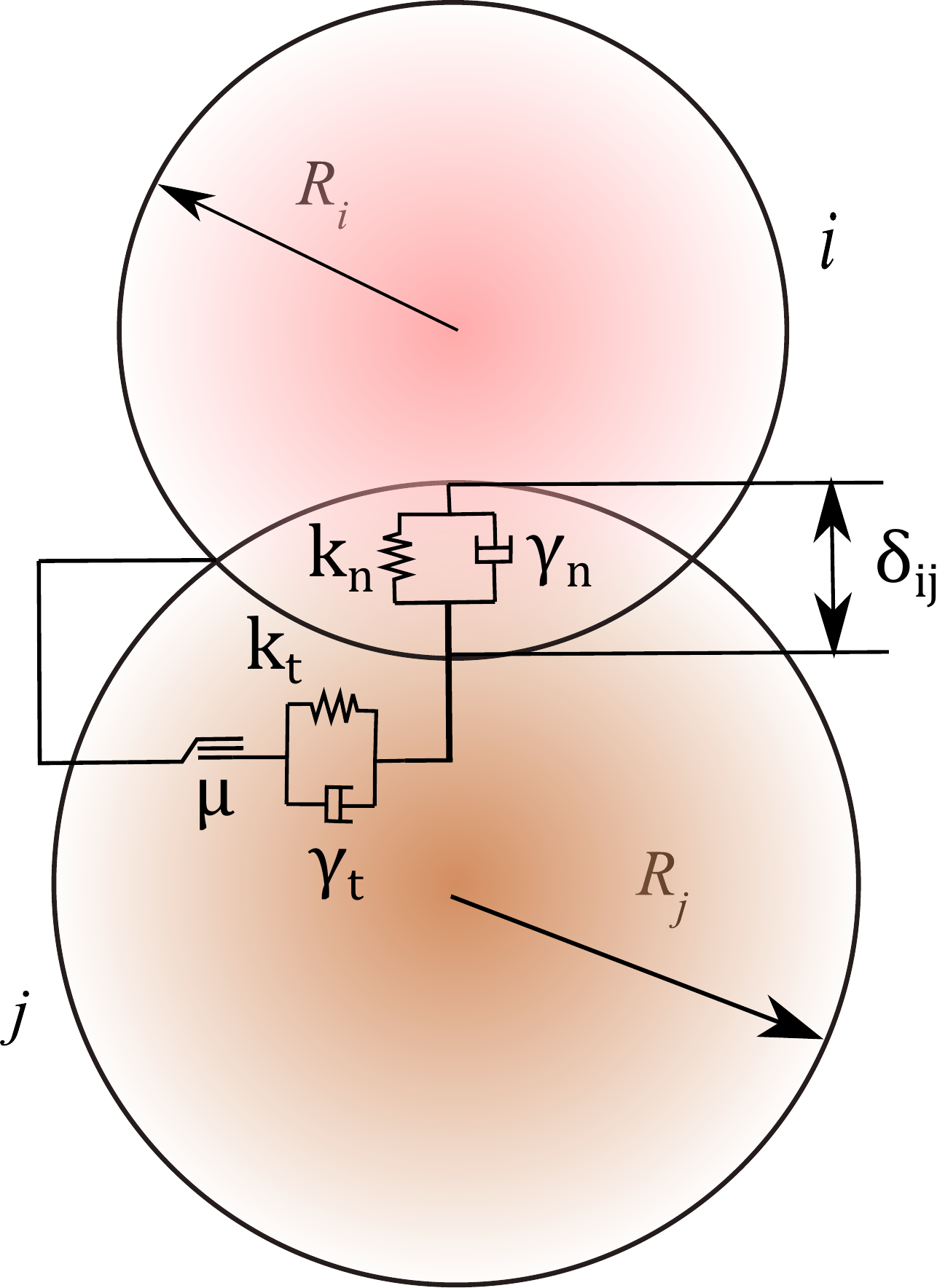}
		\caption{Spring-dashpot-slider model for the interaction forces between two deformable soft grains during collision in (a) the normal direction and (b) the tangential direction.}
		\label{fig:LSD}
	\end{center}
\end{figure}

In the tangential direction the interaction force is 

\begin{equation}
\mathbf{F}_{ij}^t=\left\{
  \begin{array}{@{}ll@{}}
    -k_t\delta_t\mathbf{t} - \xi_t\mathbf{v}_t, & \text{if}\ \left\vert\frac{F_{ij}^t}{F_{ij}^n}\right\vert < \mu_s\\
    -\mu_d \mid \mathbf{F}_{ij}^n\mid \mathbf{t}, & \text{otherwise}
  \end{array}\right.
\end{equation}

where $k_t$, $\xi_t$, and $\mu$ are respectively the tangential spring stiffness coefficient, the tangential damping coefficient, and the friction coefficient of the slider in the tangential direction.

\subsection{Geometry of the system}
The geometry consists of a horizontal cylindrical barrel of diameter $D_b$ inside which is a screw placed concentrically. The diameter of the screw-shaft is $D_s$ and the clearance between the screw-blade and the barrel is $\mathcal{O}(d_p/10)$. The length of the screw conveyor considered for the study is equal to one pitch length of the screw. Simulations were done under two broad categories. In the first category, the barrel size was kept constant and the pitch set to integral multiples of the barrel diameter. In the second category, the pitch was held constant and the barrel size was adjusted so as to achieve desired gaps between the screw-shaft and the barrel.

\begin{table}
	\centering
	\setlength{\tabcolsep}{10pt} 
	\renewcommand{\arraystretch}{2} 
\begin{tabular}{l|c}
	\hline
	\textbf{Geometric parameters} & - \\ 
	\hline \hline
	screw-shaft diameter ($D_s$) & $13 \, d_p$ \\
	\hline 
	barrel diameter ($D_b$) &$38 \, d_p$ \\
	\hline
	pitch ($p$)&\begin{tabular}{c}$38 \, d_p$, $76 \, d_p$, $114 \, d_p$, \\ $152 \, d_p$, $190 \, d_p$\end{tabular} \\
	\hline
	blade thickness ($t_b$)&$1 \, d_p$\\
	\hline
\end{tabular}
\caption{Dimensions of the geometric parameters involved in the DEM setup}
\end{table}

\FloatBarrier
\subsection{Selection of model parameters}
The Discrete Element Method with linear spring-dashpot-slider
(LSDS) contact model has been used to compute the motion of grains. The open source software, LIGGGHTS\textsuperscript{\textregistered} \citep{kloss2012models} which is an enhanced version of 
Large-scale Atomic/Molecular Massively Parallel Simulator (LAMMPS\textsuperscript{\textregistered}) \citep{plimpton1995fast} package
has been employed. LIGGGHTS\textsuperscript{\textregistered} enables us to import complex geometries into the simulation domain as triangulated mesh files. The values of the parameters used in the simulation are listed in Table \ref{tab:parameters} 

\begin{table*}
\centering
\setlength{\tabcolsep}{10pt} 
\renewcommand{\arraystretch}{2} 
\begin{tabular}{l | c | c }
\textbf{Quantity} & \textbf{Formula} & \textbf{Value}  \\
\hline \hline
particle diameter$(d_p)$ &-& 1 $mm$ \\
density$(\rho)$ &-& 2500 $kg/m^3$ \\
mass$(m)$ & $\rho \frac{ \pi d_p^3}{6}$ & $1.31*10^{-6} \, kg$ \\
normal spring constant$(k_n)$ &$10^6 \frac{mg}{d_p}$& $12834.75 \, N/m$ \\
tangential spring constant$(k_t)$ &$\frac{2}{7}k_n$& $3667.07 \, N/m$\\
normal damping constant$(\gamma_n)$ &$317 m \sqrt{\frac{g}{d_p}}$& $0.04$ \\
tangential damping constant$(\gamma_t)$ &$\frac{1}{2} \gamma_n$& $0.021$\\
collision time$(t_c)$ &$\pi (\frac{2k_n}{m}-\frac{\gamma_n^2}{4m^2})^{-1/2}$& $2.26*10^{-5} \,s$ \\
simulation timestep$(\Delta t)$ &$\frac{t_c}{10}$& $2.26*10^{-6} \,s$
\end{tabular}
\caption{Model Parameters}
\label{tab:parameters}
\end{table*}

\FloatBarrier
\subsection{Overview of the simulation process}
Firstly, the desired mixture of particles is poured into the system through the hole cut out in the barrel. The number distribution used in the current simulation is 30\%, 40\% and 30\% of 0.9 $d_p$, 1.0 $d_p$ and 1.1 $d_p$ respectively. A distribution in the particle size reduces the probability of crystallization of the system during operation. After the system has been filled upto the desired height, the pouring stops and the barrel is replaced with its no-hole counterpart. Gravity
can be turned off at this point and the screw is set to rotate at a constant rpm. The rotation continues for 2-3 complete turns after which the kinematics and dynamics of the flow are studied.\\
\begin{figure}
	\begin{minipage}{.5\linewidth}
		\begin{subfigure}{\linewidth}
			\centering
			\includegraphics[width=.9\linewidth]{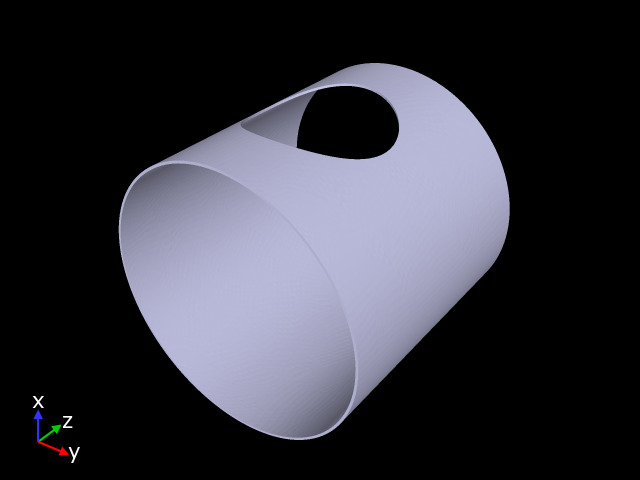}
			\caption{}
			\label{fig:sub1}
		\end{subfigure}\\[1ex]
		\begin{subfigure}{\linewidth}
			\centering
			\includegraphics[width=.9\linewidth]{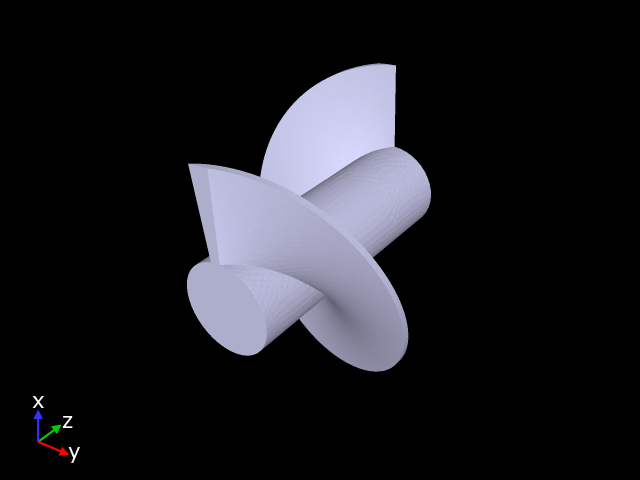}
			\caption{}
			\label{fig:sub2}
		\end{subfigure}
	\end{minipage}%
	\begin{minipage}{.5\linewidth}
		\begin{subfigure}{\linewidth}
			\centering
			\includegraphics[width=.9\linewidth]{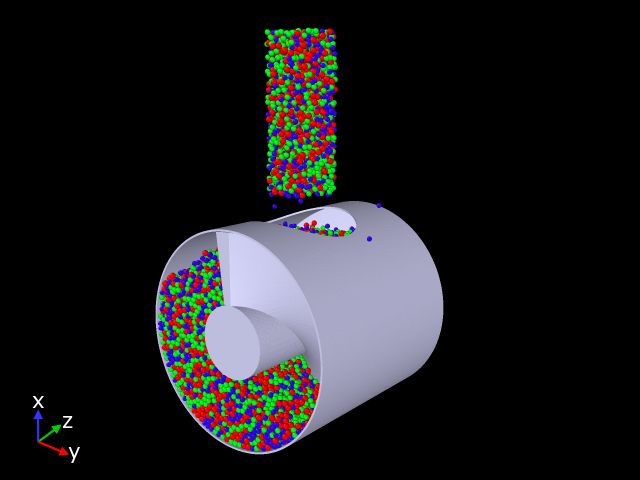}
			\caption{}
			\label{fig:sub4}
		\end{subfigure}\\[1ex]
		\begin{subfigure}{\linewidth}
			\centering
			\includegraphics[width=.9\linewidth]{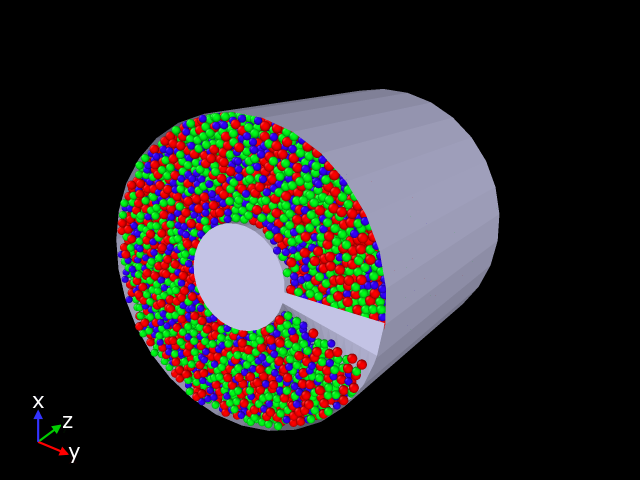}
			\caption{}
			\label{fig:sub5}
		\end{subfigure}
	\end{minipage}
	\caption{Components in the simulation. (a) barrel that encapsulates the screw with a temporarily cut hole in it (b) single pitch of the screw (c) particles being poured into the conveyor through the hole, under gravity (d) barrel replaced with its no-hole counterpart and the screw set to rotation}
	\label{fig:test}
\end{figure}
\FloatBarrier

\subsection{The post-processing analysis}
To compute the continuum variables from the position, velocity and force on each particle, we divide the region of interest into bins. Over each bin, averaging of properties of interest is carried out.
However, we wait until the fluctuation in the total kinetic energy of the system becomes negligible.

\begin{figure}
	\centering
	\includegraphics[width=0.6\linewidth]{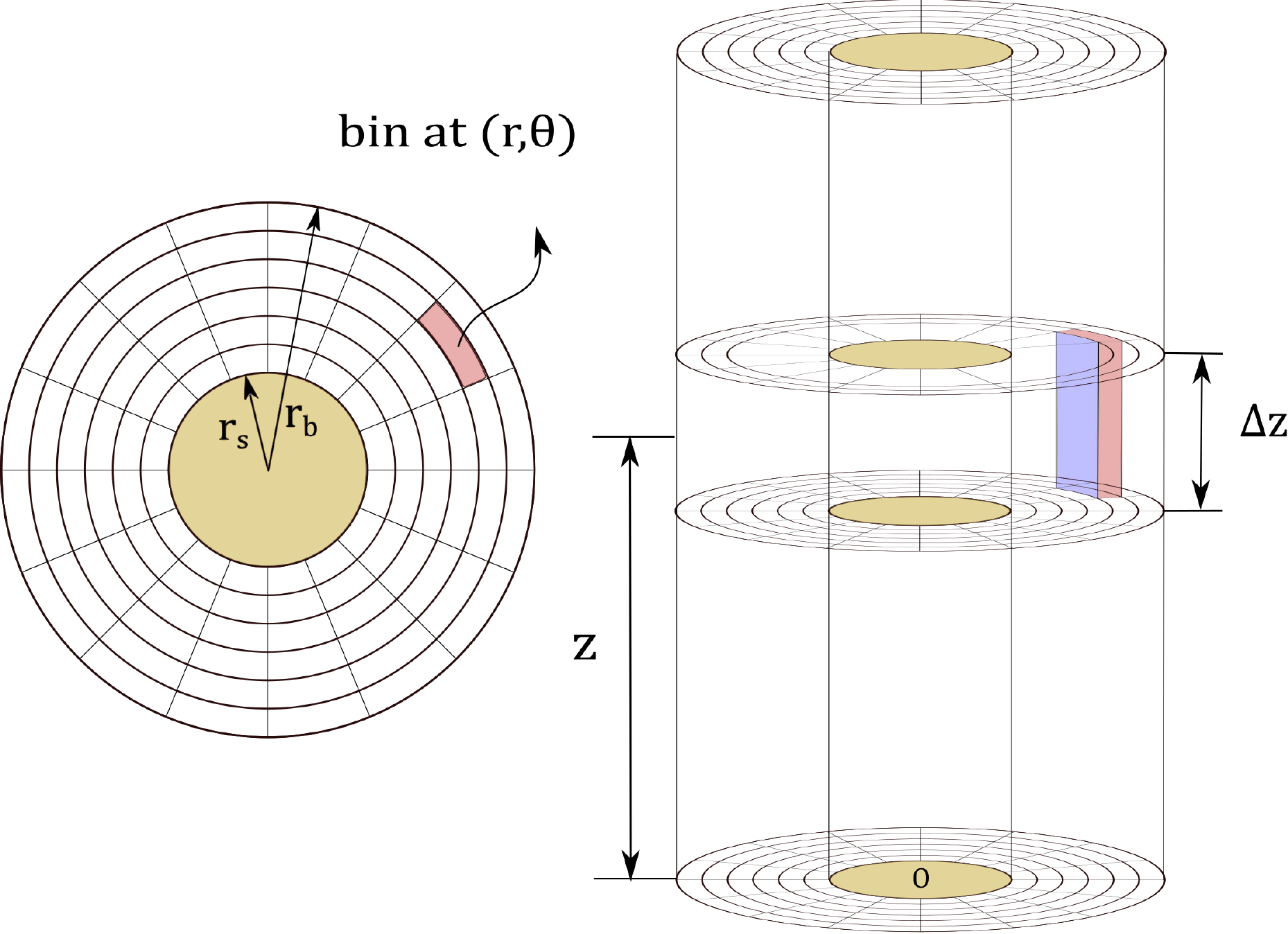}
	\caption{Segmentation of the simulation domain into bins}
	\label{binning}
\end{figure}

In the current work, a bin looks similar to what is depicted in Figure \ref{binning}. Consider a plane perpendicular to the axis of the screw at a distance $z$ from the origin $O$. The portion of the plane that lies outside the screw-shaft and inside the barrel is then extruded by a distance $\frac{\Delta z}{2}$ in either direction. The volume is then partitioned with cylindrical shells whose radii go from $r_s$ to $r_b$ in equal steps. The region is further segmented by a set of planes generated by rotating a plane that accommodates the axis of the screw about the axis itself, in equal angular increments. 

Recall that there is the rotating screw flight in the annular region which is constantly cutting through several bins. However, the bins are fixed in time and space and what we report is the flux of grains averaged over integral number of screw rotations at each bin location. As we are simulating a fully developed system, we can further average the results from bins at different $z$ locations that have the same $r$ and $\theta$ components.

\end{document}